\documentclass[12pt]{article}
\usepackage{comment}
\usepackage{textcomp}
\usepackage{chetnew}
\usepackage{amssymb,amsmath,amsfonts,mathrsfs,braket}
\usepackage[utf8]{inputenc}
\usepackage{graphicx,longtable,afterpage}

\usepackage{floatrow}
\newfloatcommand{capbtabbox}{table}[][\FBwidth]

\def\one{{\,\hbox{1\kern-.8mm l}}}

\newcommand{\CC}{\mathcal{C}}

\allowdisplaybreaks

\def\makeatletter{\catcode`\@=11}
\makeatletter
\def\mathbox#1{\hbox{$\m@th#1$}}%
\def\math@ccstyles#1#2#3#4#5#6#7{{\leavevmode
      \setbox0\mathbox{#6#7}%
      \setbox2\mathbox{#4#5}%
      \dimen@ #3%
      \baselineskip\z@\lineskiplimit#1\lineskip\z@
      \vbox{\ialign{##\crcr
             \hfil \kern #2\box2 \hfil\crcr
             \noalign{\kern\dimen@}%
             \hfil\box0\hfil\crcr}}}}
\def\mathaccstyles{\math@ccstyles\maxdimen}
\def\maththroughstyles{\math@ccstyles{-\maxdimen}}
\def\unity%
 {\maththroughstyles{.45\ht0}\z@\displaystyle {\mathchar"006C}\displaystyle 1}

\def\CC{{\cal C}}

\def\GG{{\cal L}}  

\def\NN{{\cal N}}
\def\OO{{\cal O}}

\def\SS{{\cal S}}

\def\beq{\begin{equation}}
\def\eeq{\end{equation}}
\newcommand{\bea}{\begin{eqnarray}}
\newcommand{\eea}{\end{eqnarray}}
\def\bal{\begin{align}}
\def\eal{\end{align}}

\preprint{CCTP-2023-5 \hfill QMUL-PH-23-10 \\ ITCP-IPP-2023/5}

\title{\vspace{-1.cm} 
Bootstrability in Line-Defect CFT \\ \vspace{0.3cm} 
with Improved Truncation Methods} 

\author{
V.~Niarchos\;$^{a,\bigstar}$, C.~Papageorgakis\;$^{b,\diamondsuit}$, P.~Richmond\;$^{b,\spadesuit}$, A.~G.~Stapleton$^{b,\heartsuit}$, M.~Woolley$^{b,\clubsuit}$}

\affiliation{
$^a$ ITCP \& CCTP, Department of Physics,\\
University of Crete, 71003 Heraklion, Greece\\
$^b$ Centre for Theoretical Physics, Department of Physics and Astronomy\\ Queen Mary University of London, London E1 4NS, UK \vspace{0.3cm} $ $ \\
\vspace{0.3cm} $ $

\vspace{0.3cm}
{\tt \small
$^\bigstar$niarchos@physics.uoc.gr,  
$^\diamondsuit$c.papageorgakis@qmul.ac.uk, 
$^\spadesuit$p.richmond@qmul.ac.uk,
$^\heartsuit$a.g.stapleton@qmul.ac.uk,
$^\clubsuit$mitchell.woolley@qmul.ac.uk}}

\abstract{We study the conformal bootstrap of 1D CFTs on the straight Maldacena--Wilson line in 4D ${\cal N}=4$ super-Yang--Mills theory. We introduce an improved truncation scheme with an `OPE tail' approximation and use it to reproduce the `bootstrability' results of Cavagli\`a et al.\ for the OPE-coefficients squared of the first three unprotected operators. For example, for the first OPE-coefficient squared at 't Hooft coupling $(4\pi)^2$, linear-functional methods with two sum rules from integrated correlators give the rigorous result $0.294014873 \pm 4.88 \cdot 10^{-8}$, whereas our methods give with machine-precision computations $0.294014228 \pm 6.77 \cdot 10^{-7}$. For our numerical searches, we benchmark the Reinforcement Learning Soft Actor-Critic algorithm against an Interior Point Method algorithm (IPOPT) and comment on the merits of each algorithm.}

\date{\today}

\begin{document}

\maketitle

\hypersetup{pageanchor=true}

\setcounter{tocdepth}{2}

\toc

\section{Introduction and Summary}
\label{summary}

In two recent papers, the authors of Refs.\ \cite{Cavaglia:2021bnz,Cavaglia:2022qpg} initiated a non-perturbative study of  1D defect Conformal Field Theories (CFTs) in the planar 't Hooft limit, combining methods from integrability and the numerical conformal bootstrap programme. The analysis of \cite{Cavaglia:2021bnz,Cavaglia:2022qpg} focused on the 1D defect CFT of the $\frac{1}{2}$-BPS infinite Maldacena--Wilson line in 4D $\NN=4$ super-Yang Mills (SYM) theory. It produced high-precision numerics for three different (non-supersymmetric) three-point functions involving two protected and one unprotected operator. To obtain these results, the usual constraints of crossing symmetry were combined with information about spectral data from integrability\footnote{The input of exact spectral data into the conformal bootstrap has been considered for 2D CFTs in \cite{Picco:2016ilr,He:2020rfk}.} and two sum rules arising from integrated correlation functions \cite{Drukker:2022pxk,Cavaglia:2022yvv}.

The combination of powerful techniques from integrability and the conformal bootstrap, dubbed {\it bootstrability} in \cite{Cavaglia:2021bnz}, aims to blend two methods that have played a leading role in non-perturbative studies of Quantum Field Theories. On the one hand, integrability has proven very successful in the analytic computation of scaling dimensions in the planar limit of gauge theories, but less efficient in computations of correlation functions. On the other hand, the conformal bootstrap\footnote{For reviews see \cite{Simmons-Duffin:2016gjk,Poland:2018epd,Chester:2019wfx}, while for a recent state-of-the art \cite{Hartman:2022zik,Poland:2022qrs}.} has shown great promise in yielding rigorous results for generic CFT data (including both scaling dimensions and correlation functions), but has difficulty in navigating towards arbitrary theories of interest. The input of integrability guides a conformal bootstrap search towards a desired solution.

In this work, we continue the bootstrability study of the 1D CFT on the straight, $\frac{1}{2}$-BPS Maldacena--Wilson line, by employing a different methodology on the conformal bootstrap side compared to Refs.\ \cite{Cavaglia:2021bnz,Cavaglia:2022qpg}. Instead of the commonly used {\it linear functional} method \cite{Rattazzi:2008pe}, which produces rigorous inequalities for CFT data by making use of the positivity constraints from unitarity, we introduce an {\it improved truncation scheme} to directly solve approximate crossing equations and sum rules. 

Truncation methods within the conformal bootstrap programme are not new, having previously yielded outcomes with varying degrees of success. Earlier work includes \cite{Gliozzi:2013ysa,Gliozzi:2014jsa,Gliozzi:2015qsa,Gliozzi:2016cmg,Esterlis:2016psv,Hikami:2017hwv,Hikami:2017sbg,Li:2017ukc}, while more recently  \cite{Kantor:2021jpz,Kantor:2021kbx,Kantor:2022epi} and \cite{Poland:2023vpn} have implemented truncation methods in the context of the four- and five-point bootstrap respectively. In particular, references \cite{Kantor:2021jpz,Kantor:2021kbx} highlighted the significance of selecting an appropriate truncation strategy as a means of guiding the search within the conformal bootstrap framework. These studies also aimed to devise methodologies capable of handling substantial truncations, encompassing hundreds or even thousands of operators. This represents a notable improvement over earlier works, which employed more drastic truncations limited to $O(10)$ operators. 

One of the main disadvantages of truncation methods is that they are subject to systematic errors that are hard to quantify, rendering them non-rigorous. Furthermore, determining the appropriate truncation can be a non-trivial task, and the interpretation of results obtained through such approaches may not always be straightforward. Conversely, truncation methods have advantages such as flexibility and computational cost-effectiveness. They do not rely on positivity constraints, making them well-suited for exploring the landscape of CFT data, including the cases of non-unitary theories, defect CFTs and analyses arising from bootstrapping higher-point functions. Therefore, truncation-based searches could be creatively employed in guiding targeted searches for specific theories, extracting dynamically viable gap assumptions and other information that a more rigorous method could employ at a later stage. We find this aspect particularly interesting and worthy of further exploration.

In this paper, we expand upon the bootstrability investigation of the 1D defect CFT but also introduce several improvements to previously-employed truncation methods. These improvements allow us to accurately reproduce the findings of \cite{Cavaglia:2021bnz,Cavaglia:2022qpg} with exceptional precision. This precision extends to the seventh decimal place for certain data, all the way from the weak to the strong coupling regimes. To obtain these results we employed both Reinforcement Learning (as done in \cite{Kantor:2021jpz, Kantor:2021kbx, Kantor:2022epi}), as well as more conventional non-linear optimisation algorithms. We used the same information from integrability methods (the scaling dimensions of 10 non-protected operators) as developed in \cite{Grabner:2020nis, Julius:2021uka} and employed for bootstrability in \cite{Cavaglia:2022qpg}.

\begin{table}[t!]
\centering
\begin{tabular}{|c | c | c | c|} 
 \hline
 Method & $C_1^2$ & $C_2^2$ & $C_3^2$ \\ [0.5ex] 
 \hline\hline
 {\small SDPB in \cite{Cavaglia:2022qpg}}  & {\small 0.294014873 $\pm$ 4.88$\cdot 10^{-8}$} & \small{0.039788 $\pm$ 4.10$\cdot 10^{-4}$} & \small{0.146757 $\pm$ 5.82$\cdot 10^{-4}$}  \\ 
 \hline
 {\small Improved Truncation}  & {\small 0.294014228 $\pm$ 6.77$\cdot 10^{-7}$} & {\small 0.041832 $\pm$ 1.86$\cdot 10^{-3}$} & {\small 0.144100 $\pm$ 2.39$\cdot 10^{-3}$} \\ [1ex] 
 \hline
\end{tabular}
\caption{\footnotesize{Sample results for three OPE-coefficients squared in the 1D line-defect CFT of the Maldacena--Wilson line in planar 4D $\NN=4$ SYM theory at 't Hooft coupling $\lambda = (4\pi)^2\simeq 157.91$ or $g=1$ in the notation of the main text. The precise definitions appear in Section~\ref{1dcft}. The first line presents results from \cite{Cavaglia:2022qpg} using the linear-functional methods implemented with the Semidefinite Programming approach of SDPB \cite{Simmons-Duffin:2015qma} and the errors reflect rigorous upper/lower bounds. The second line presents a sample of our new results based on the use of an improved truncation method that employed an Interior Point optimisation algorithm implemented with IPOPT \cite{Wachter}. The errors are statistical in this case and reflect $1\sigma$ deviation. In the main text we present further results obtained using alternative optimisers. Both lines combine the crossing equations with two sum rules from integrated correlation functions. The full list of results is available in Table~\ref{tab:resall}.}}
\label{Tab_specimen}
\end{table}

Our main results can be summarised as follows:
\begin{itemize}
    \item We present a substantially improved truncation scheme by introducing approximations for the `tail' of the OPE expansion, as well as `effective' operators.
    \item We obtain numerical results for the OPE-coefficients squared of the first three unprotected multiplets of the 1D line-defect CFT, with/without utilising the sum rules of integrated correlators and compare with \cite{Cavaglia:2021bnz,Cavaglia:2022qpg}. A sample of these results is presented in Table~\ref{Tab_specimen}, with the full list available in Table~\ref{tab:resall}.

    \item We benchmark the Soft Actor-Critic (SAC) Reinforcement Learning algorithm \cite{DBLP:journals/corr/abs-1801-01290}, used as a non-linear optimiser, against the IPOPT implementation of the Interior Point Method optimisation algorithm. In this context, we provide specific evidence for the effectiveness of the average of statistical runs with the SAC algorithm and comment on the motivation to explore more advanced Reinforcement Learning algorithms. We highlight the fact that truncation methods using either of the above algorithms are computationally cheaper than the (rigorous) linear-functional methods. Our computations were performed using machine precision.
    \item We significantly improve our Python implementation \href{https://github.com/vniarchos/BootSTOP}{BootSTOP} to: $a)$ use the improved truncation scheme in 1D, $b)$ work with CFTs in 1D, 2D and 6D, $c)$ switch between SAC and the Python Parallel Global Multiobjective Optimiser (PyGMO) \cite{Biscani2020}, which includes a host of deterministic and stochastic optimisation algorithms, including IPOPT. 
    
\end{itemize}

The rest of this paper is organised as follows. In Section~\ref{truncation} we give a detailed account of our improved truncation scheme. In Section~\ref{1dcft} we provide a brief review of the necessary background needed to set up our bootstrap problem. In Sections~\ref{wointegral} and \ref{wintegral} we list our main results by reproducing the recently obtained values for the OPE-coefficients squared of the first three unprotected multiplets in the 1D defect CFT \cite{Cavaglia:2021bnz, Cavaglia:2022qpg} using our SAC/IPOPT optimisation protocols. We round off in Section~\ref{comparison} by discussing the merits of different optimisation algorithms, as well as presenting some predictions for additional data in the 1D defect CFT, before concluding in Section~\ref{outlook}. Appendix~\ref{fullresults} includes a full list of our results, while Appendix~\ref{flow} presents an alternative analysis for the adiabatic variation of the crossing equations, which could be used in future studies.\footnote{Note added in v2: After v1 of this work appeared on the arXiv, we learned of similar explorations of the 1D defect CFT using BootSTOP in the master's thesis \cite{TrentaMSc}. We thank P. Ferrero for communication on this point.}

\section{Improved Truncation Schemes}
\label{truncation}

One of our objectives in this paper is to investigate enhancements to truncation methods. Specifically, we aim to develop a novel framework for interpreting the operators and associated CFT data within a truncation scheme. We will attempt a systematic treatment in contexts where the CFT is a member of a parametric family (with continuous or discrete parameters), by assuming that the theory can be defined/solved in a specific corner of that family. Let us call this corner of parameter space the {\it defining corner}. That corner typically reflects a weak coupling formulation and could be a free fixed point, or a generalised free point (possibly captured by a dual supergravity description). The known spectrum of the defining corner forms the starting point for an informed truncation of the spectrum. The goal of the programme is to explore how the CFT data evolve (adiabatically) across the parameter space.\footnote{A recent study of the Ising CFT using a different adiabatic deformation in spacetime dimension appeared in \cite{Henriksson:2022gpa}.} This strategy fits well within a more general approach that prioritises the exploration of specific theories, compared to an exploration of general properties in the space of CFTs. 

The ensuing discussion will be kept generic and applies to CFTs in any number of spacetime dimensions. For concreteness, we will focus on a single crossing equation, but similar methods can also be applied to the multi-correlator bootstrap, or to the bootstrap with the addition of extra sum rules. The typical bootstrap problem involves an algebraic crossing equation of the general form
\begin{equation}
    \label{truncaa}
    \sum_{n} \CC_n F_n(x) + r(x) = 0~,
\end{equation}
where the index $n$ runs over the infinite number of operators that appear in the conformal block expansion (in multiple channels). In two and higher spacetime dimensions, $n$ enumerates operators of different spin and scaling dimension. In one dimension, where there is no concept of spin, $n$ simply labels operators at different scaling dimensions. $\CC_n$ denotes the OPE-coefficients squared and $F_n$ is shorthand notation for the (crossed) conformal blocks. The variable $x$ represents the single cross ratio present in 1D CFTs or collectively the pair of complex cross-ratios $(z,\bar z)$ in higher-dimensional CFTs. The function $r(x)$ is an in-homogeneous contribution, which is assumed to be explicitly known.\footnote{In the upcoming Eq.\ \eqref{eq:7}, $r(x)=\mathcal{G}_{\rm simple}(g,x)$.} This term may depend on external continuous or discrete parameters. The CFT data encoded in \eqref{truncaa} are the OPE-coefficients squared $\CC_n$ and the corresponding scaling dimensions $\Delta_n$.

The cross-ratio dependence of the crossing equation can be discretised, either by evaluating it on a grid of $x$-points \cite{CastedoEcheverri:2016fxt} or by applying a finite number of linear functionals. A popular basis of linear functional in the conformal bootstrap literature consists of derivatives at the crossing-symmetric point \cite{Simmons-Duffin:2016gjk,Poland:2018epd,Chester:2019wfx}. For the 1D applications of the upcoming sections we used (even) derivatives at $x=\frac{1}{2}$. This discretisation reduces the continuous character of the algebraic equation \eqref{truncaa} to a finite subset of equations, which we collect in a finite-dimensional vector. Accordingly, we recast Eq.\ \eqref{truncaa} into the vector form:
\begin{equation}
    \label{truncab}
    \sum_{n} \CC_n \vec{F}_n + \vec{r} = 0~.
\end{equation}

Let us now split the full set of operators appearing in \eqref{truncab} into a finite subset, call it $\SS$, and its complement. The selection of $\SS$ can be based on various criteria, which we do not have to specify at the moment. Typically, we are interested in a subset of `the most significant' operators. In the Euclidean bootstrap around the crossing symmetric point, where the conformal block expansion converges exponentially fast \cite{Pappadopulo:2012jk}, these are operators with relatively low scaling dimensions.\footnote{Higher-dimension operators also play a significant role in our approach and how we incorporate them into $\SS$ is part of our discussion. Summarising remarks related to this aspect appear in Section\ \ref{comparison}. More generally, we expect that higher-dimension operators will eventually become increasingly important in hybrid numerical/analytical bootstrap methods; see e.g.\ \cite{Su:2022xnj} for a recent discussion.} Consequently, we can now recast Eq.\ \eqref{truncab} into the more refined form
\begin{equation}
    \label{truncac}
    \sum_{n\in \SS} \CC_n \vec{F}_n + \vec{T} + \vec{r} = 0~,
\end{equation}
where $\vec T$ captures the contribution of the operators in the {\it complement} of $\SS$, which we will call the `tail', and the sum over $n$ now involves only a finite number of terms and corresponding CFT data. Thus far, \eqref{truncac} is exact. 

The main premise of truncation methods, up to this point in the literature, involves dropping the tail contribution $\vec T$ completely and analysing the resulting equation, 
\begin{align}\label{eq:vec}
    \vec E := \sum_{n\in \SS} \CC_n \vec{F}_n + \vec{r} \simeq 0\;,
\end{align} 
which can only be satisfied approximately. In \cite{Gliozzi:2013ysa} the analysis of the truncated equations proceeds via the method of determinants. In \cite{Li:2017ukc}, and subsequently in \cite{Kantor:2021kbx,Kantor:2021jpz}, one formulates a positive semi-definite function $\GG$ of the vector $\vec E$ and tries to minimise the `cost' function
\begin{equation}
    \label{truncad}
    \mathrm{Cost}\left[\{\Delta_n, \CC_n\}_{n\in \SS}\right] := \GG[\vec E]
    ~.
\end{equation}
$\GG$ quantifies the deviation of $\vec E$ from the zero vector, and should therefore vanish at zero by definition. A typical choice of $\GG$ is the root mean square but other options can also be explored. 

We want to depart, slightly, from this logic by keeping the tail $\vec T$ with a suitable approximation, and obtaining a better understanding of what the data $\{\Delta_n, \CC_n\}_{n\in \SS}$ represent in an approximate scheme, with or without $\vec T$. Part of the problem relates to the fact that in truncations with many operators the higher-dimension CFT data can be redistributed by the optimisation algorithm in many different ways, to collectively capture a similar overall, approximate contribution to the cost function. This includes configurations where operators are grouped together in narrow bands of scaling dimensions, effectively reducing the number of active, independent CFT data in the truncation. There is also an interplay between this freedom and the dynamics of the tail, that affects the complexity of the optimisation problem and the interpretation of the results for a given truncation scheme.

\subsection{Tail Approximation}
\label{tail1}
As  mentioned in the beginning of this section, we will assume that the theory of interest is part of a family of CFTs, and that there is at least one corner in parameter space where it can be solved explicitly with traditional methods. In order to set up a convenient language, let us collectively denote the external parameters $\lambda$, and their value at the defining corner $\lambda^*$. The parameters $\lambda$ could be discrete (e.g.\ the rank of a gauge group) or continuous (e.g.\ the value of an exactly marginal coupling). In general, the notation $\lambda$ is shorthand for a multi-parameter vector. 

The existence of an explicit solution at $\lambda^*$ has several useful consequences. First, the solution at $\lambda^*$ can be used to inform the choice of truncation, namely the set of operators $\SS$ in the crossing equation \eqref{truncac}. For instance, this can be done by picking a cutoff on the scaling dimension or twist, so as to specify the number of operators that we want to include in $\SS$ for each spin. In \cite{Kantor:2021jpz,Kantor:2021kbx}, this choice informed a corresponding  `spin-partition'. With the number of data appearing in $\SS$ specified, our goal takes the following form:

\vspace{0.3cm}
\noindent
{\it Solve \eqref{truncac} to determine how the data $\{\Delta_n, \CC_n\}_{n\in \SS}$ vary across the parameter space from $\lambda^*$ to a generic value of $\lambda$.}
\vspace{0.3cm}

The quality of the results can depend non-trivially on the tail $\vec T$, which includes the value of the CFT data of all the hidden operators in the complement of $\SS$. As a first step towards a better approximation of $\vec T$ (compared to simply setting $\vec T=0$) we propose the following approach: At $\lambda^*$, we assume having access to the CFT data $\{\Delta_n^*, \CC_n^*\}_{n\in \SS}$ inside the truncation, and the equation \eqref{truncac} can be satisfied exactly:
\begin{equation}
    \label{tail1aa}
    \sum_{n\in \SS} \CC_n^* \vec{F}_n^* + \vec{T}^* + \vec{r}^* = 0
    ~.
\end{equation}
Most importantly, we can use this equation to determine the exact value of the tail at $\lambda^*$:
\begin{equation}
    \label{tail1ab}
    \vec{T}^* = - \sum_{n\in \SS} \CC_n^* \vec{F}_n^* - \vec{r}^*
    ~.
\end{equation}
If the tail $\vec T$ does not vary significantly as a function of $\lambda$, then a first approximation of the tail consists of setting
\begin{equation}
    \label{tail1ac}
    \vec T(\lambda) \simeq \vec T^*
    ~.
\end{equation}
In such a case the exact equation \eqref{truncac} is approximated by
\begin{equation}
    \label{tail1ad}
    \sum_{n\in \SS} \CC_n \vec{F}_n + \vec{r} - \sum_{n\in \SS} \CC_n^* \vec{F}_n^* - \vec{r}^* \simeq 0
\end{equation}
and leads to the minimisation of the modified cost function
\begin{equation}
    \label{tail1ae}
    \widetilde{\mathrm{Cost}}\left[\{\Delta_n, \CC_n\}_{n\in \SS}\right] := \GG[\vec E - \vec E^*]
    ~,
\end{equation}
with  $\vec E$ defined in \eqref{eq:vec}.

The approximate assumption in \eqref{tail1ac} is not unrealistic (for sufficiently large truncations) in the vicinity of $\lambda^*$ and certainly improves the drastic truncation ansatz $\vec T=0$. Indeed, there is now at least one point in parameter space where the crossing equations are satisfied exactly by construction. The assumption \eqref{tail1ac} is also motivated by the fact that high-dimension operators have minimal contribution to the conformal block expansion around the crossing-symmetric point, and that in the limit of high spin, CFT states behave asymptotically as generalised free fields \cite{Komargodski:2012ek,Fitzpatrick:2012yx}. Nevertheless, whether this approximation holds for a finite deformation away from $\lambda^*$ (and to what degree) is not obvious and is certainly critical. In general, one can imagine various ways in which \eqref{tail1ac} can break down. For example, as one deforms away from $\lambda^*$ and the spectrum rearranges itself, some operators from the tail can become increasingly important. The tail contribution can also be affected when the scaling dimensions of the external operators are $\lambda$-dependent.\footnote{ For example, one can explicitly write down the crossing equations in the 2D $S^1$ CFT for scalar, charged primaries \cite{Kantor:2021jpz,Kantor:2021kbx} and check the value of the tail for a fixed truncation as a function of the external operator dimensions. As one moves on the conformal manifold, the value of the external dimension changes and the tail exhibits significant variations. We would like to thank A.~Stratoudakis for working out specific examples of this type.}

\subsection{Effective Operators and Degeneracies}
\label{degeneracies}

Setting the approximation of the tail aside for the moment, another issue that affects the complexity and efficiency of a truncation scheme relates to the presence of large accidental degeneracies. This usually occurs in the defining corner at $\lambda^*$ that involves a (generalised) free field description; the weak coupling regimes of gauge theories are typical examples. Large degeneracies are challenging for two reasons: First, they can grow very rapidly as functions of the scaling dimension. In that case, a complete description of the degenerate spectrum would force a truncation with high dimensionality. Second, away from $\lambda^*$ the accidental degeneracies are typically lifted, and tracking the precise splitting across the parameter space can be a very complicated task. One might therefore ask: Is it possible to alleviate the problems that arise in such situations?

To isolate the effects of nearly-degenerate operators, let us assume that part of the sum $\sum_{n\in \SS} \CC_n \vec{F}_n$ in \eqref{tail1ad} involves a relatively narrow band of $\mathfrak N_{\tt band}$ operators (at the same spin) with scaling dimensions $\Delta \in {\tt band}$, where ${\tt band} \equiv [\Delta_{\min}, \Delta_{\max}]$ and $\Delta_{\max} > \Delta_{\min}$. We will denote their contribution to the crossing equation as
\begin{equation}
    \label{degaa}
    \vec{\mathfrak E} = \sum_{\Delta \in {\tt band} \subset \SS} \CC_n \vec{F}_n
    ~.
\end{equation}
For an exact solution to the crossing equations this vector takes a specific value
\begin{equation}
    \label{degab}
    \vec{\mathfrak E}^{\rm(exact)} = \sum_{\Delta \in {\tt band} \subset \SS} \CC_n^{\rm(exact)} \vec{F}_n^{\rm(exact)}
    ~.
\end{equation}
We want to explore the possibility of approximating the exact vector $\vec{\mathfrak E}^{\rm(exact)}$ with an effective sum 
\begin{equation}
    \label{degac}
    \vec{\mathfrak E}^{(\rm eff)} = \sum_{\OO_{\rm eff}} \CC_{\OO_{\rm eff}} \vec{F}_{\OO_{\rm eff}}
\end{equation}
over a reduced number $\mathfrak N_{\tt eff}$ of operators. Crucially, the CFT data of these operators do not capture the exact data of the CFT in the band. They are meant to provide an effective description that approximates the contribution $\vec{\mathfrak E}^{\rm(exact)}$ inside the crossing equation.

A special instance where this effective description is exact is that of exact degeneracies. In that case, there may be a possibly large number of distinct operators, $\mathfrak N_{\tt band}>1$, that contribute to the sum $\vec{\mathfrak E}^{\rm(exact)}$ in \eqref{degab}. However, since all of them have the same scaling dimension $\Delta$ (and the same corresponding conformal block $\vec{F}_{\Delta}^{\rm(exact)}$), the vector $\vec{\mathfrak E}^{\rm(exact)}$ is effectively encoding the contribution of a single operator $(\mathfrak N_{\tt eff}=1)$ with OPE-coefficient squared equal to the sum of the OPE-coefficients squared  of the individual degenerate operators:
\begin{equation}
    \label{degad}
    \vec{\mathfrak E}^{\rm(exact)} = \left(\sum_{\Delta_n = \Delta} \CC_n^{\rm(exact)}\right) \vec{F}_{\Delta}^{\rm(exact)}
    ~.
\end{equation}
Therefore, from this single effective operator only the scaling dimension $\Delta$ and  total OPE-coefficient squared  $\left(\sum_{\Delta_n = \Delta} \CC_n^{\rm(exact)}\right)$ can be read off.

More generally, in a band of finite size one can write 
\begin{equation}
    \label{degae}
    \vec{\mathfrak E} = \sum_{\Delta \in {\tt band}} \CC_n \vec{F}_n = \bar \CC \sum_{\Delta \in {\tt band}} c_n \vec{F}_n
    ~,
\end{equation}
where we defined
\begin{equation}
    \label{degaf}
    \bar \CC := \sum_{\Delta \in {\tt band}} \CC_n~, ~~ 
    c_n := \frac{\CC_n}{\bar \CC}
    ~.
\end{equation}
With this definition, and assuming that $\CC_n\geq 0$ by unitarity, the new coefficients $c_n$ are by construction numbers inside the interval $[0,1]$ with the property $\sum_n c_n = 1$. Consequently, the vector $\vec{\mathfrak e}^{(\rm exact)} = \sum_{\Delta \in {\tt band}} c_n^{(\rm exact)} \vec{F}_n^{(\rm exact)}$ of the exact solution is inside the convex combination ${\bf \Sigma}_{(\rm exact)}$ of the $\mathfrak N_{\tt band}$ vectors $\vec F_n^{(\rm exact)}$. Moreover, the convex combination ${\bf \Sigma}_{(\rm exact)}$ is inside the convex hull ${\boldsymbol{H}}[\Delta_{\min}, \Delta_{\max}]$ of the segment of the curve $\vec F_{\Delta}$ for $\Delta \in [\Delta_{\min}, \Delta_{\max}]$. The latter is a set that characterises the conformal blocks independently of the details of the exact solution of the $\mathfrak N_{\tt band}$ operators in the band. To summarise, the exact contribution of the band to the crossing equation is the vector
\begin{equation}
    \label{degag}
    \vec{\mathfrak E}^{(\rm exact)} = \bar \CC\, \vec{\mathfrak e}^{(\rm exact)}
\end{equation}
with
\begin{equation}
    \label{degai}
    \vec{\mathfrak e}^{(\rm exact)} \in {\bf \Sigma}_{(\rm exact)} \subset {\boldsymbol{H}}[\Delta_{\min}, \Delta_{\max}]
    ~.
\end{equation}

When the $\mathfrak N_{\tt band}$ operators are replaced by $\mathfrak N_{\tt eff} < \mathfrak N_{\tt band}$ operators, the quality of the approximation will depend on the minimal distance between the convex combination ${\bf \Sigma}_{(\rm eff)}$ of the $\mathfrak N_{\tt eff}$ vectors $\vec F_n^{(\rm eff)}$ and the convex combination ${\bf \Sigma}_{(\rm exact)}$ as the scaling dimensions of the effective operators vary. Assuming the latter vary inside the same band $[\Delta_{\min}, \Delta_{\max}]$ as the scaling dimensions of the exact configuration, both convex combinations ${\bf \Sigma}_{(\rm eff)}$ and ${\bf \Sigma}_{(\rm exact)}$ are subsets of the same convex hull ${\boldsymbol{H}}[\Delta_{\min}, \Delta_{\max}]$. This puts an indirect upper bound on the error of the approximation of the exact configuration.

It is not easy to promote these observations into specific quantitative predictions in generic situations, or to use them to develop a concrete strategy for the selection of the effective operators. We wanted, however, to highlight these features for two reasons. 

First, we believe that an effective description of a complicated spectrum can be an important tool that can be used to reduce the complexity of the problem. For relatively narrow bands of nearly-degenerate operators one might expect reasonable results with cheap effective descriptions. Moreover, parameterising ignorance with an effective description may lead to a better interpretation of the output of a computation. For instance, if there is confidence in the existence of a nearly-degenerate band for a given problem, then instead of trying to interpret specific numbers as individual predictions for actual CFT data, it may be more appropriate to interpret those results as features of an effective description. In that case, from the spread of the scaling dimensions of the effective operators one may want to distil a prediction for the size of the band, and from the overall coefficient $\bar \CC$ in \eqref{degae} one may want to distil an approximate sum rule for the total OPE-coefficient squared in the band.

A second related motivation for the above discussion is that sometimes, during the optimisation steps in a high-dimensional truncation, an algorithm (or two separate algorithms) may identify two distinct high-reward configurations with one of them having rendered several operators nearly-degenerate. Rather than interpreting these two configurations as results corresponding to two distinct theories with a different number of operators, it may be more appropriate to view them as different effective representations of the same theory.

\subsection{Implementation and a Soft Extension of the Tail}
\label{softtail}

In the last subsection we attempted to isolate effects inside some relatively narrow band of operators. Let us now return to the complete problem and the approximate truncation scheme \eqref{tail1ad}. 

In the defining corner at $\lambda^*$, we understand the structure of the spectrum and how it is captured by our chosen truncation. As we deform the theory away from $\lambda^*$ we can now envision the emergence of the following complications: nearly-degenerate bands (possibly captured by a reduced set of effective operators) can develop significant splits, operator scaling dimensions can cross and the naive approximation of the tail at $\lambda^*$ may cease to be accurate. The latter will force the operators inside our truncation set $\SS$ to readjust appropriately. How one proceeds at this point depends on the situation, and will typically require additional external input in order to extract confident results. For example, such an input could arise by considering the simultaneous information from multiple correlators, the combination of a truncation scheme with a navigator method based on the linear-functional approach \cite{Reehorst:2021ykw} and/or input from OPE inversion formulae \cite{Caron-Huot:2017vep, Simmons-Duffin:2017nub,Carmi:2019cub}. We plan to explore all these possibilities in future work. In the present paper, the external input that we use are the exact scaling dimensions for 10 operators from the Quantum Spectral Curve. 

Accordingly, our results in Sections\ \ref{wointegral}, \ref{wintegral} are obtained with a truncation of 62 operators, further split into 10 operators, the scaling dimensions of which we can track explicitly, and the remaining 52 operators that we treat as effective. We will not attempt to make any predictions for actual CFT data based on these effective operators. We sum their contribution to the crossing equation and treat it as part of a soft extension of the initial tail approximation $\vec T^*$ at zero 't Hooft coupling. We will provide concrete evidence that this soft extension of the tail is a valid approximation at all values of the coupling. We will also see that different algorithms treat the 52 effective operators in different ways.

\section{Review of Defect CFT in $\NN=4$ SYM}
\label{1dcft}

Before delving into the details of our numerical computations, we begin with a lightning summary of the 1D line-defect CFT, highlighting only the aspects that are necessary for our discussion. For a complete account we refer the reader to \cite{Cavaglia:2022yvv} and references therein. The line-defect CFT resides on a straight,  infinite Maldacena--Wilson line
\begin{align}
  \label{eq:1}
  \mathcal W = \mathrm{Tr~ P ~exp}\int_{- \infty}^{+\infty}(A_t + \Phi_{||}) dt
\end{align}
that preserves an $\mathfrak{osp}(4^*|4)$ subalgebra of the full superconformal algebra of the parent $\mathcal N=4$ super Yang--Mills theory in four dimensions \cite{Drukker:2006xg}, and inherits its integrable structure in the planar limit \cite{Drukker:2012de,Correa:2012hh}. In \eqref{eq:1} $A_t$ and $\Phi_{||}$ are gauge and real scalar-field components respectively. The maximal bosonic subalgebra involves the 1D conformal algebra, the $\mathfrak{sp}(2)_R$ R-symmetry and the algebra of $\mathfrak{so}(3)$ rotations transverse to the line-defect in four-dimensional spacetime (sometimes referred to as `spin'). All states in the line CFT fall into unitary irreducible representations of this superconformal algebra, the superconformal primaries of which are scalars under the $\mathfrak{so}(3)$ global symmetry. The irreducible representations include  short $\mathcal B_k$ (protected) representations, the dimension of which is fixed by their $\mathfrak{sp}(2)_R$ quantum numbers $[0,k]$, $\Delta = k $,  and long $\mathcal L^\Delta_{[0,0]}$ (unprotected) representations which are $\mathfrak{sp}(2)_R$ scalars. 

We are interested in four-point functions arising from local-operator insertions along the Maldacena--Wilson line. More specifically, we are interested in identical insertions of one of the real scalars of $\mathcal N= 4$ SYM $\Phi_{\perp}^i$, $i = 1, \ldots, 5$, not appearing in \eqref{eq:1}:
\begin{align}
  \label{eq:2}
  \langle \langle \Phi^1_\perp(t_1) \Phi^1_\perp(t_2) \Phi^1_\perp(t_3) \Phi^1_\perp(t_4)\rangle\rangle :=  \langle \mathrm{Tr} W_{-\infty}^{t_1} \Phi^1_\perp(t_1) W_{t_1}^{t_2} \Phi^1_\perp(t_2) W_{t_2}^{t_3}\Phi^1_\perp(t_3) W_{t_3}^{t_4} \Phi^1_\perp(t_4) W_{t_4}^{+\infty}\rangle
  ~.
\end{align}
The $\Phi^1$ component is the superconformal primary of the $\mathcal B_1$ multiplet, known as the displacement multiplet, the OPEs of which obey the following selection rules:
\begin{align}\label{eq:3}
    \mathcal{B}_1 \times \mathcal{B}_1=\mathcal{I}+\mathcal{B}_2+\sum_{\Delta>1} \mathcal{L}_{[0,0]}^{\Delta}\;.
\end{align}

A crossing equation arises from \eqref{eq:2} due to the invariance of the four-point function  under a cyclic relabelling of the insertion points, which can be recast as
\begin{align}\label{eq:6}
    x^2 f(1-x)+(1-x)^2 f(x)=0\;.
\end{align}
In this expression, $x$ is the single conformal cross-ratio in 1D,
\begin{align}
    x := \frac{x_{12} x_{34}}{x_{13} x_{24}}, \quad x_{i j} := x_i-x_j\;.
\end{align}
As a consequence of \eqref{eq:3} the function $f(x)$ admits a superconformal-block decomposition of the form
\begin{align}\label{eq:4}
    f(x)=F_{\mathcal{I}}(x)+C_{\mathrm{BPS}}^2 F_{\mathcal{B}_2}(x)+\sum_n C_n^2 F_{\Delta_n}(x)\;,
\end{align}
with the specific blocks given by
\begin{align}
F_{\mathcal{I}}(x) & =x \\
F_{\mathcal{B}_2}(x) & =x-x_2 F_1(1,2,4 ; x) \\
\label{eq:5} F_{\Delta_n}(x) & =\frac{x^{\Delta_n+1}}{1-\Delta_n}{ }_2 F_1(\Delta_n+1, \Delta_n+2,2 \Delta_n+4 ; x)
\end{align}
involving standard hypergeometric functions. 

In addition to fixing the dimension of the $\mathcal B_2$ primary from superconformal representation theory for all values of the 't Hooft coupling $\lambda$ of $\mathcal N = 4 $ SYM, one can also determine the value of the corresponding OPE-coefficients squared, $C_{\mathrm{BPS}}^2$, with the help of supersymmetric localisation \cite{Giombi:2017cqn, Liendo:2018ukf} or integrability methods \cite{Cavaglia:2022qpg}. The latter vary with the 't Hooft coupling and, when expressed as a function of $g := \frac{\sqrt \lambda}{4 \pi}$, read
\begin{align}
    C_{\mathrm{BPS}}^2 = 1- \mathbb F(g)\;,
\end{align}
where
\begin{align}
    \mathbb{F}(g)=\frac{3\left(g^2-\mathbb{B}(g)\right)}{\pi^2(\mathbb{B}(g))^2}
\end{align}
and the Bremsstrahlung function is
\begin{align}
\mathbb{B}(g)=\frac{g}{\pi} \frac{I_2(4 \pi g)}{I_1(4 \pi g)}\;,
\end{align}
involving modified Bessel functions of the first kind. 

Using this information, the crossing equations \eqref{eq:6} can be recast into the following compact form
\begin{align}\label{eq:7}
    \sum_n C_n^2 \mathcal{G}_{\Delta_n}(x)+\mathcal{G}_{\text {simple }}(g, x)=0\;,
\end{align}
where 
\begin{align}
    \mathcal{G}_{\text {simple }}(g, x) := \mathcal{G}_{\mathcal{I}}(x)+C_{B P S}^2(g) \mathcal{G}_{\mathcal{B}_2}(x)
\end{align}
is a now a known function, with $\mathcal G_{\mathcal I}$ encoding the crossed superconformal blocks:
\begin{align}
    \mathcal{G}_{\mathcal{I}, \mathcal{B}_2, \Delta_n }(x) := (1-x)^2 F_{\mathcal{I}, \mathcal{B}_2, \Delta_n }(x)+x^2 F_{\mathcal{I}, \mathcal{B}_2, \Delta_n }(1-x)\;.
\end{align}
Therefore, the undetermined quantities appearing in the superconformal-block expansion of \eqref{eq:7} will involve the dimensions of the unprotected operators along with their corresponding OPE-coefficients squared. The lowest dimension long primary is $\Phi_{||}$. The long CFT data vary as a function of $g$ but one can make use of the Quantum Spectral Curve (QSC) to numerically determine the evolution of the conformal dimensions of long operators from weak to strong coupling. These results arise from a 1D adaptation of a numerical QSC implementation for $\mathcal N = 4$ SYM developed in \cite{Gromov:2015wca,Gromov:2023hzc}. The dimensions of the first 35 long operators for $g\in [0,2]$ were provided in \cite{Cavaglia:2021bnz}, while those of the first 13 long operators for $g \in [0,4]$ were found using methods developed in \cite{Grabner:2020nis, Julius:2021uka} and used in \cite{Cavaglia:2022qpg}.

In \cite{Cavaglia:2021bnz} the dimensions of the first 2 unprotected superconformal primaries were used as external input to the linear-functional method to determine bounds for the OPE coefficient of the first long multiplet entering \eqref{eq:7}. Preliminary results with the input of additional scaling dimensions from the QSC were also reported. In \cite{Cavaglia:2022qpg}, the dimensions of only the first 10 unprotected superconformal primaries were used but the crossing symmetry conditions were supplemented by two sum rules arising from integrated correlators, which were derived in \cite{Drukker:2022pxk} and \cite{Cavaglia:2022yvv} respectively. The incorporation of these sum rules was observed to yield significantly sharper bounds and better accuracy for the first three OPE-coefficients squared. These two additional constraints from integrated correlators can be brought to the following form:
\begin{align}\label{integral}
\sum_{\Delta_n} C_n^2 \operatorname{Int}_i\left[F_{\Delta_n}(x)\right]+\mathrm{RHS}_i=0 \qquad \textrm{for} \qquad i = 1,2\;,
\end{align}
where 
\begin{align}
& \operatorname{Int}_1[F_{\Delta_n}(x)] := -\int_0^{\frac{1}{2}}\left(x-1-x^2\right) \frac{F_{\Delta_n}(x)}{x^2} \partial_x \log (x(1-x)) d x \;,\\
& \operatorname{Int}_2[F_{\Delta_n}(x)] := \int_0^{\frac{1}{2}} F_{\Delta_n}(x)\frac{(2 x-1)}{x^2}dx\;,
\end{align}
with $F_{\Delta_n}(x)$ given in \eqref{eq:5} and
\begin{align}
\mathrm{RHS}_1 & =\frac{\mathbb{B}-3 \mathbb{C}}{8 \mathbb{B}^2}+\left(7 \log (2)-\frac{41}{8}\right)(\mathbb{F}-1)+\log (2)\;, \\
\mathrm{RHS}_2 & =\frac{1-\mathbb{F}}{6}+(2-\mathbb{F}) \log (2)+1-\frac{\mathbb{C}}{4 \mathbb{B}^2}\;.
\end{align}
The curvature function $\mathbb C(g)$ has analytic expansions at weak and strong coupling, while a numerical evaluation with high precision, which we used in our implementation, was provided in \cite{Cavaglia:2022yvv}.

\section{Results without Integral Constraints}
\label{wointegral}

We now move on to present the main results of this paper. We begin with the analysis of the crossing symmetry conditions \eqref{eq:7} without any input from the two integral constraints \eqref{integral}. We will fix the scaling dimensions of the first 10 long multiplets using the QSC and compare with the corresponding results in \cite{Cavaglia:2022qpg}, which implemented the linear-functional method with SDPB. The inclusion of the integral constraints will be discussed separately in Section~\ref{wintegral}.

Our goal is to extract information about the scaling dimensions and corresponding OPE-coefficients squared for operators in unprotected (long) multiplets that appear in the conformal block expansion of the four-point function \eqref{eq:2}. An adiabatic sequence of runs was performed on the HPC cluster at Queen Mary University of London (QMUL) starting at $g = 0.2$ up to $g=4$ with step $\delta g=0.2$. We analysed derivatives of the crossing equations at the crossing symmetric point (as per usual practice in the conformal bootstrap programme) and chose to normalise each derivative with a factor of $1/(2^p \, p!)$ at order $p$. Because of symmetry, only the even derivatives are non-trivial. In most of the reported results we included all even derivatives up to order $N_{der} = 260$, but we also performed runs with $N_{der}=700$.

\subsection{Implementation of Algorithms}

In this section, we will report independent results using two optimisation methods. The first is based on the Soft Actor-Critic (SAC) algorithm, deployed as a stochastic optimiser and implemented through PyTorch. This is a Reinforcement Learning algorithm based on the concept of Markov Decision Processes, first introduced in this context in \cite{Kantor:2021kbx,Kantor:2021jpz}.\footnote{There have been many recent applications of Machine Learning techniques to high-energy theoretical  physics. An incomplete list of references includes: the exploration of string vacua \cite{Abel:2021ddu, Abel:2021rrj, Cole:2021nnt}, integrability \cite{Krippendorf:2021lee, Lal:2023dkj},  the construction of numerical Calabi--Yau metrics for string compactifications \cite{Ashmore:2019wzb,Larfors:2022nep, Berglund:2022gvm}, interplays with Wilsonian Renormalisation in Quantum Field Theory \cite{Hashimoto:2019bih,Halverson:2020trp, Erbin:2021kqf,Berman:2022uov,Berman:2023rqb}, String Field Theory \cite{Erbin:2022rgx} and lattice Quantum Field Theory \cite{deHaan:2021erb,Gerdes:2022eve}. For a recent review, see \cite{He:2023csq}. For an introduction to Reinforcement Learning and the SAC algorithm see \cite{Kantor:thesis}.} The second is the Interior Point optimisation (IPOPT) algorithm \cite{Wachter}, which is deterministic. We note that IPOPT is now accessible from within our coding framework for numerical bootstrap, \href{https://github.com/vniarchos/BootSTOP}{BootSTOP}, which incorporates all the algorithms of the Python Parallel Global Multi-objective Optimiser \href{https://github.com/esa/PyGMO2}{PyGMO} \cite{Biscani2020}.\footnote{BootSTOP currently contains libraries for conformal blocks in 1D, 2D and 6D, necessary for attacking CFTs in the corresponding spacetime dimensions with truncation methods. We intend to make further updates with 3D and 4D conformal blocks in the near future.} Access to this library allows the user to choose from a large suite of standard deterministic and stochastic algorithms. In our problem, IPOPT seems to outperform other algorithms available in PyGMO (such as Simulated Annealing, Particle Swarm Optimisation and Differential Evolution); we have not, however, performed a full, systematic, comparative study of all the PyGMO options. 

We used the highest precision possible in the PyTorch and PyGMO packages: floating point precision, which on 64-bit machines corresponds to 16 decimal places. In order to improve the runtime of the algorithms we pre-generated the values of the differentiated conformal blocks. We found a closed form expression for the $p^{\mathrm{th}}$ derivative and verified this formula in Mathematica up to order $p=20$. Beyond this order Mathematica became extremely slow so we opted to evaluate the $N_{der}=700$ derivatives in Python using the MPMATH package. Whilst this package allows for arbitrary precision we chose 20 decimal places for all intermediate calculations before reducing to 16 for the final output. Cross-checks with Mathematica were performed when this was possible. Each derivative was evaluated on a lattice of conformal weights starting from 0 and ending at 20 with an increase of $10^{-4}$ between lattice sites.  Conformal blocks on scaling dimensions in-between the points of the grid were evaluated with linear interpolation. The pre-generation of conformal blocks, and our setup within PyTorch and PyGMO, are some of the main obstacles towards arbitrary numerical precision in our implementation, and the reason why we restricted our computations to machine precision. It is encouraging that this compromise did not significantly affect the quality of our final results.

Both SAC and IPOPT algorithms were employed to minimise the $L^2$-norm cost function of the difference between the value of the crossing equations \eqref{eq:7} at each coupling $g$ and its corresponding value at $g=0$. This approach implements the improved truncation scheme of Section~\ref{truncation} with the tail approximated by its value at $g=0$, as set up in \eqref{tail1ae}. We will also be frequently referring to the corresponding `reward' of a configuration, defined as the inverse of the $L^2$-norm cost function. 

\subsection{Choice of Truncation}

\begin{table}
\centering
 {\small   \begin{tabular}{|c|c|c|c|c|c|c|c|c|c|c|}
 \hline
$J$ & 1 & 2 & 3 & 4 & 5 & 6 & 7 & 8 & 9 & 10 \\
 \hline\hline
$\#$ of operators in truncation & 1 & 2 & 6 & 22 & 8 & 8 & 5 & 4 & 3 & 3  \\
 \hline
\end{tabular}}
\caption{\footnotesize{The number of operators included in our truncation. The operators are allocated in groups characterised by the $g=0$ value of their scaling dimension, $J$. For the low-lying operators at $J=1,2,3$ this exactly matches the known $g=0$ degeneracy. For $J\geq 4$ the number of chosen operators is smaller than the expected $g=0$ degeneracy. We used more operators at $J=4$ (3 above the reported 19 operators in \cite{Cavaglia:2021bnz,Cavaglia:2022qpg}) and gradually less at higher values of $J$. All the operators in the groups $J\geq 4$ are therefore effective. In total, our truncation involves 62 operators with 124 corresponding CFT data.}}
\label{tab:partition}
\end{table}

The set $\SS$ of long operators that we included in this truncation was informed by the structure of the spectrum at $g=0$. At this free point, the scaling dimensions of long operators are integer, $\Delta\equiv J=1,2,\ldots$. With the exception of $J=1$, all other levels are degenerate, with degeneracies that can be determined in principle. Our choice is detailed in Table \ref{tab:partition}. The operators were grouped according to their $g=0$ scaling dimension, $J$. Only the first 9 states at $J=1,2,3$ match the exact degeneracy at $g=0$. All the states at higher values of $J$ are effective in the sense of Section~\ref{degeneracies}.

As the coupling $g$ is increased, the  spectrum rearranges itself and the free-theory degeneracies are lifted. At each search cycle, we opted to reorder the operators within the same $J$ family according to their scaling dimension, as we would normally do in higher-dimensional CFTs for the tower of states at each spin \cite{Kantor:2022epi}. Such a  choice allowed us to track how the groups of nearly-degenerate states evolved with increasing coupling. However, we note that $J$ is not a spin quantum number. All the operators within the truncation contribute with the same type of conformal block and at higher values of $g$ the mixing between different $J$ sectors is significant. This is a characteristic difficulty of 1D CFTs that does not exist in higher dimensions.

 Therefore, in all our runs we included 62 operators in the truncation, which amounts to 124 CFT data (62 scaling dimensions and 62 corresponding OPE-coefficients squared), although technically our numerical algorithms can also handle efficiently many more operators. Input from the QSC was used for the scaling dimensions of the first 9 operators at $J=1,2,3$ and for the lowest one at $J=4$. Our main results refer to the OPE-coefficients squared of the $J=1,2$ operators, denoted respectively as $C_1^2, C_2^2, C_3^2$.

\subsection{Specifics of the SAC Runs}

The SAC searches were implemented with $N_{der}=260$ derivative constraints. We employed 200 agents, each allowed to run on the QMUL HPC cluster for a maximum of 23 hours.\footnote{We observed that most agents were approaching their final configuration roughly within the first 12 hours. We did not attempt to optimise the scheduling of the algorithm, opting to allow for a longer search. In BootSTOP, SAC was implemented with parameters: ${\tt faff\_ max} = 5000$, ${\tt pc\_ max} = 6$, ${\tt window\_ rate} = 0.7$, ${\tt max\_ window\_ exp} = 30$.} The search windows ({\tt guess$\_$sizes$\_$deltas} and {\tt guess$\_$sizes$\_$opes} in the code) were set to $0.4$ for the conformal dimensions of the 52 unfixed long multiplets, $2\times 10^{-2}$ for the OPE-coefficients squared of the first 47 long multiplets and $2\times 10^{-3}$ for the OPE-coefficients squared of the last 15 long multiplets. Each run was performed around the (reward-weighted) average of the solutions at the previous value of $g$. 

In the SAC implementation, the scaling dimensions of the 10 `fixed' long multiplets were not completely fixed. They were allowed to vary with a small $10^{-3}$ search window around the solution at the previous value of $g$. We stress that in SAC this value does not control the size of the box inside which the search is performed. Instead, it controls the maximum size of the next action. In this manner, if the starting value of a datum is not in the vicinity of a local minimum, the algorithm can eventually wander off significantly, even with a small search window. 

At the end of the runs for the 200 SAC agents at each $g$, we also did an independent search with IPOPT inside a $4\sigma$ area around the average of the SAC result. Two sets of independent IPOPT runs were executed here, one with $N_{der} = 260$ and another with $N_{der} = 700$ derivative constraints. These follow-up IPOPT runs were performed with $2\times 10^8$ agents (subdivided on the cluster into 2k jobs, each with a population of 100k in the PyGMO architecture). They increased the reward significantly by a couple of orders of magnitude, but did not move the SAC averages. The addition of more derivative constraints did not lead to significant improvement either. We will make additional comments regarding these features in Section~\ref{comparison}. 

\begin{figure}[t]
\centering
\includegraphics[width=16cm]{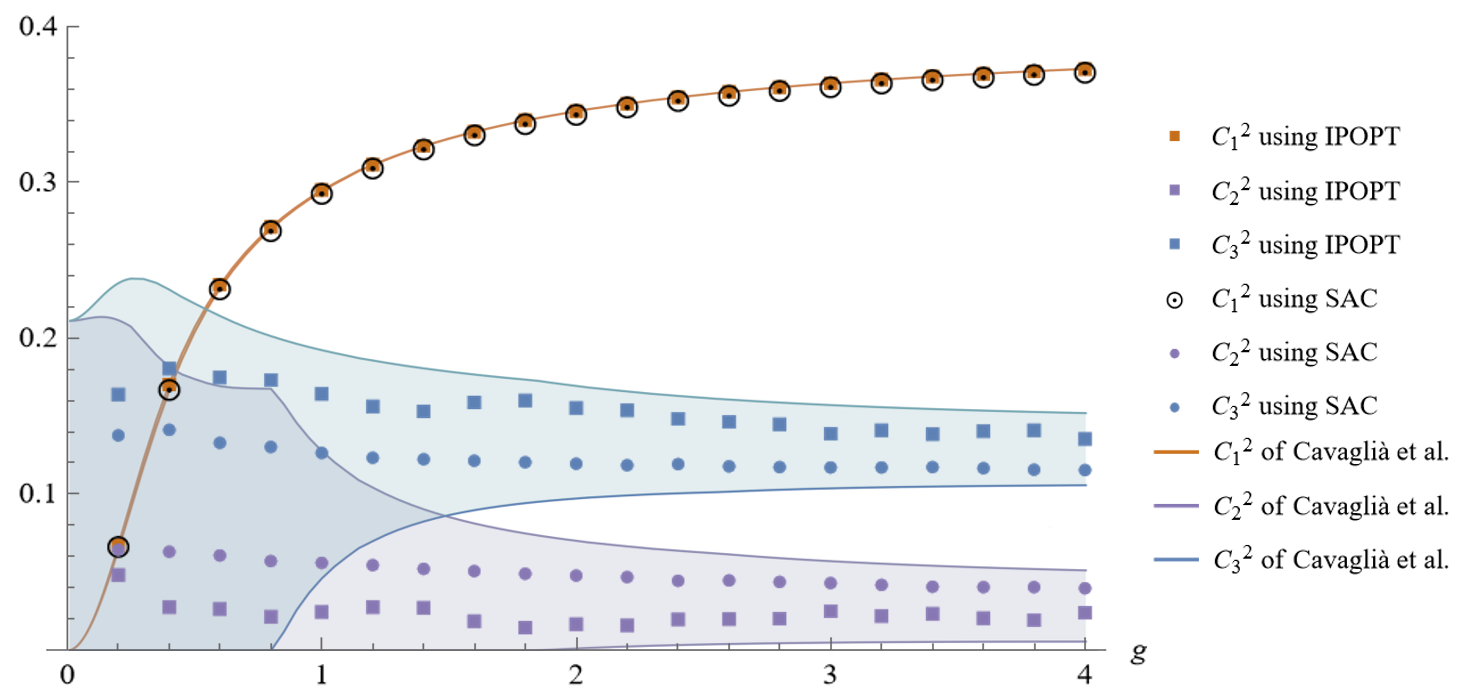}
\caption{\footnotesize{Results for the OPE-coefficients squared of the first three long operators with no integral constraints. The solid lines indicate the rigorous bounds presented in Figure 6 of \cite{Cavaglia:2021bnz}, reprinted here with permission from the authors. Same-coloured  circles and squares indicate our results from the SAC and IPOPT runs respectively. The corresponding statistical errors are too small to display on this plot but can be found in Table~\ref{tab:resall}.}}
\label{fig:nocons}
\end{figure}

\subsection{Specifics of the IPOPT Runs}

The IPOPT algorithm was employed with $4\times 10^8$ agents. These runs were subdivided into groups with a population of 100k within the PyGMO architecture. Each group was run 4k times on the QMUL HPC cluster with an approximate 20 minutes runtime. Our final statistics for this approach comprise the 200 HPC cluster runs with the highest reward. The results we report in Table \ref{tab:res1} were obtained with $N_{der} = 260$ derivative constraints. The box of the overall search was fixed within the hyperparameters of the algorithm. We chose $\pm 1$ for the scaling dimensions and $\pm 0.2$ for the OPE coefficients, around the solution at the previous value of $g$. We also enforced as extra lower bounds the free-limit value for the scaling dimension in each $J$ family, and 0 for the OPE-coefficients squared. In contrast to the SAC runs, the first 10 long operators had their dimensions completely fixed to the results of the QSC. To further assist the search, we imposed on both algorithms (SAC and IPOPT) the additional constraints $C_2^2 < 0.1$ and $C_3^2>0.1$. 

\subsection{Results}

\begin{table}[t]
\centering
 {\small   \begin{tabular}{|l|c|c|c|c|}
 \hline
Method & $g$ & $C_1^2$ & $C_2^2$ & $C_3^2$ \\
\hline\hline
\cite{Cavaglia:2021bnz}  & $0.2$ & $0.0663 \pm 1.9 \cdot 10^{-3}$ &  & \\
IPOPT & $0.2$ & $0.06607342 \pm 4.18 \cdot 10^{-5}$ & $0.04708 \pm 2.04 \cdot 10^{-3}$ & $0.1630 \pm 2.69 \cdot 10^{-3}$ \\
SAC & $0.2$ & $0.06733947 \pm 1.26 \cdot 10^{-3}$ & $0.06506 \pm 1.05 \cdot 10^{-2}$ & $0.1384 \pm 1.47 \cdot 10^{-2}$ \\
 \hline\hline
\cite{Cavaglia:2021bnz}  & $0.4$ & $0.1684 \pm 1.9 \cdot 10^{-3}$ &  &  \\
IPOPT & $0.4$ & $0.16944584 \pm 8.35 \cdot 10^{-5}$ & $0.02659 \pm 3.45 \cdot 10^{-3}$ & $0.17965 \pm 4.90 \cdot 10^{-3}$ \\
SAC & $0.4$ & $0.16824002 \pm 1.00 \cdot 10^{-3}$ & $0.06380 \pm 1.37 \cdot 10^{-2}$ & $0.14198 \pm 1.80 \cdot 10^{-2}$ \\
 \hline\hline
\cite{Cavaglia:2021bnz}  & $0.6$ & $0.2329 \pm 9 \cdot 10^{-4}$ &  &  \\
IPOPT & $0.6$ & $0.233574606 \pm 1.32 \cdot 10^{-4}$ & $0.02533 \pm 6.78 \cdot 10^{-3}$ & $0.17382 \pm 7.68 \cdot 10^{-3}$ \\
SAC & $0.6$ & $0.232721152 \pm 3.24 \cdot 10^{-4}$ & $0.06151 \pm 5.46 \cdot 10^{-3}$ & $0.13363 \pm 6.77 \cdot 10^{-3}$ \\
 \hline\hline
\cite{Cavaglia:2021bnz}  & $0.8$ & $0.2701 \pm 5 \cdot 10^{-4}$ &  &  \\
IPOPT & $0.8$ & $0.270632286 \pm 6.67 \cdot 10^{-5}$ & $0.020165 \pm 6.29 \cdot 10^{-3}$ & $0.17218 \pm 7.06 \cdot 10^{-3}$ \\
SAC & $0.8$ & $0.270121362 \pm 2.93 \cdot 10^{-4}$ & $0.05776 \pm 5.00 \cdot 10^{-3}$ & $0.13110 \pm 5.35 \cdot 10^{-3}$ \\
 \hline\hline
\cite{Cavaglia:2021bnz} & $1.0$ & $0.29388 \pm 2.7 \cdot 10^{-4}$ &  &  \\
IPOPT & $1.0$ & $0.294177967 \pm 6.79 \cdot 10^{-5}$ & $0.023344 \pm 9.64 \cdot 10^{-3}$ & $0.163302 \pm 1.04 \cdot 10^{-2}$ \\
SAC & $1.0$ & $0.293941106 \pm 3.03 \cdot 10^{-4}$ & $0.05658 \pm 5.81 \cdot 10^{-3}$ & $0.127135 \pm 6.26 \cdot 10^{-3}$ \\
 \hline
\end{tabular}}
\caption{\footnotesize{Partial list of results (for $g\in [0.2,1]$) from IPOPT and SAC runs with no integral constraints imposed. The errors in our results encode one standard deviation around the statistical reward-weighted average. For quick reference, we have also included at each value of $g$ the results for $C_1^2$ from Ref.\ \cite{Cavaglia:2021bnz}. In that case, the errors are rigorous and have a distinctly different meaning.}}
\label{tab:res1}
\end{table}

A sample of the results obtained with SAC and IPOPT (from $g=0.2$ to $g=1$ with step $\delta g = 0.2$) appears in Table \ref{tab:res1}. The full list of results can be found in Table \ref{tab:resall} of Appendix \ref{fullresults}. In Figure\ \ref{fig:nocons} we plot the full set of results against the background of Figure\ 6 from Ref.\ \cite{Cavaglia:2022qpg}, which contains the rigorous upper/lower bounds derived with the linear-functional method and SDPB. The averages and statistical errors of the CFT data from the SAC and IPOPT runs are defined as averages weighted by the square of the ratio of current reward to the best reward.

Furthermore, in the first data row for each value of $g$ in Table \ref{tab:res1}, we have also included for quick comparison the results for $C_1^2$ from Ref.\ \cite{Cavaglia:2021bnz}. These were obtained using Algorithm 1 in Ref.\ \cite{Cavaglia:2022qpg}, which employed only 2 scaling dimensions from the QSC. Incorporating the QSC data of 10 scaling dimensions with Algorithm 2 in Ref.\ \cite{Cavaglia:2022qpg} yields slightly improved upper/lower bounds. 

The most characteristic features of our results are the following:
\begin{itemize}
    \item[(1)] The results for the OPE-coefficient squared $C_1^2$ are directly comparable with the corresponding results in Refs.~\cite{Cavaglia:2021bnz, Cavaglia:2022qpg}, with agreement at the level of the third decimal point or higher.
    \item[(2)] As is apparent from Figure\ \ref{fig:nocons}, our results for the OPE-coefficients squared $C_2^2$ and $C_3^2$ are always well within the bounds of \cite{Cavaglia:2022qpg}. Interestingly, SAC and IPOPT have not produced the same average configurations exploiting the effective operators in different ways. Towards strong coupling, the spread between the SAC and IPOPT averages seems to be an indirect probe of the rough size of the rigorous allowed regions obtained with the linear-functional method. 
    \item[(3)] The linear-functional method can be quite sensitive to the choice of spectral data imported from the QSC. It was observed in \cite{Cavaglia:2022qpg} that their algorithms ceased to converge if the spectrum deviates significantly from the QSC answer. For example, at $g = 3$ it is sufficient to introduce an error of the order of $5 \times 10^{-7}$ in the spectrum, for the method that sets bounds for $C_1^2$ to no longer converge \cite{Cavaglia:2022qpg}. In light of this, the fact that the SAC runs did not alter the values of the `fixed' conformal dimensions of the first 10 operators (even with the relatively large $10^{-3}$ window) and converged on a result with good accuracy showcases that the method is robust and managed to locate the theory. For an example of the variation in the scaling dimensions $\Delta_1, \Delta_2, \Delta_3$ in the SAC runs see Table \ref{tab:ipopt_on_sac} below.
    \item[(4)] SAC and IPOPT are producing configurations of comparable rewards.\footnote{Strictly speaking, for SAC this is true after running IPOPT around the SAC average configuration. As we discuss in more detail in Section \ref{comparison}, this has minuscule effects on the average configuration.} Qualitatively, the SAC curve in Figure~\ref{fig:nocons} is smoother compared to the IPOPT curve, but the numbers in Table \ref{tab:resall} do not declare a clear winner. Besides $C_1^2$, a more specific datum that one can check is the sum of the $C_2^2$ and $C_3^2$ coefficients. Towards the strong coupling region the scaling dimensions of the $J=2$ operators, $\Delta_2$ and $\Delta_3$, converge towards 4. As a result, the two operators remain nearly-degenerate throughout the flow from weak to strong coupling. This feature complicates the search, as was already noted in \cite{Cavaglia:2021bnz}. In Figure~4 of \cite{Cavaglia:2021bnz}, narrow bounds were reported at $g=1$ that place $C_2^2$ and $C_3^2$ on the line $C_3^2 + 1.13 C_2^2=0.19$. By inserting the average results from the SAC and IPOPT runs into this expression we find:
    \begin{equation}
        \label{sum_check_1}
        {\rm SAC}:~~ C_3^2 + 1.13 C_2^2=0.19107~, ~~~ {\rm IPOPT}:~~ C_3^2 + 1.13 C_2^2=0.18968
        ~.
    \end{equation}
    In the upcoming section we will see that the most accurate results of \cite{Cavaglia:2022qpg}, based on also using the integral constraints, yield $C_3^2 + 1.13 C_2^2=0.19171$.
\end{itemize}

\section{Results with Integral Constraints}
\label{wintegral}

\begin{figure}[t]
\centering
\includegraphics[width=16cm]{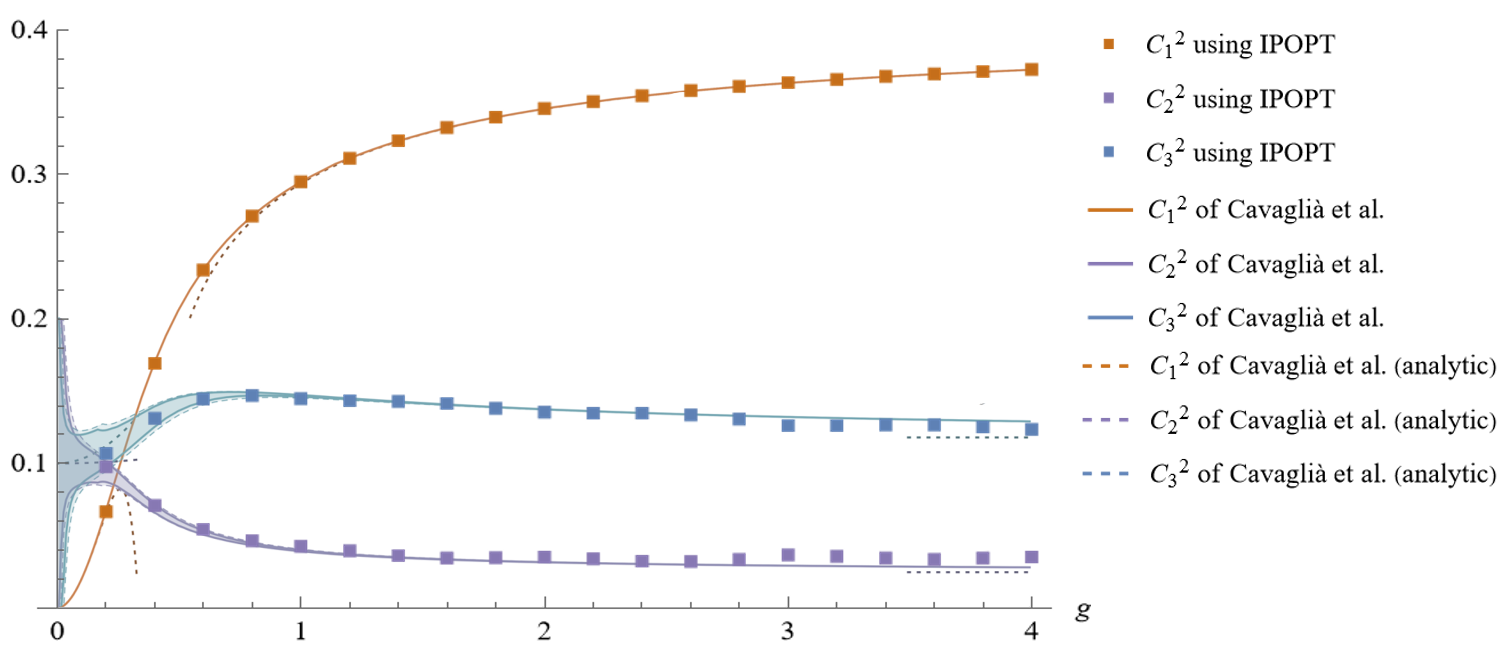}
\caption{\footnotesize{Results for the OPE-coefficients squared of the first three long operators after the incorporation of two integral constraints. The solid lines indicate the rigorous bounds presented in Figure 10 of \cite{Cavaglia:2022qpg}, reprinted here with permission from the authors. Same-colored squares indicate our results from the IPOPT runs. The corresponding statistical errors are too small to display on this plot but can be found in Table~\ref{tab:resall}.}}
\label{fig:cons}
\end{figure}

We proceed to discuss the results obtained by incorporating the two integral constraints \eqref{integral}. Anticipating a more pronounced minimum in this case, we exclusively employed IPOPT. The search parameters closely resembled those used in the IPOPT runs of Section~\ref{wintegral}. The integral constraints were supplied as separate equations using the corresponding functionality of PyGMO. Our runs involved 
$4\times 10^8$ agents. These runs were subdivided into groups with a population of 100k within the PyGMO architecture. Each group was run 4k times on the QMUL HPC cluster, with an approximate runtime of 20 minutes. Statistics were collected from the 200 runs with the highest reward. We imposed $N_{der} = 260$ derivative constraints, but, unlike the previous section, did not enforce the additional restrictions $C_2^2<0.1$ and $C_2^3>0.1$. The search windows were set to $\pm 1$ for the scaling dimensions and $\pm 0.2$ for the OPE coefficients, centred around the average solution obtained at the previous value of $g$. To ensure proper results, we incorporated lower bounds. The lower bound for each $J$ family was set to the free-limit value for the scaling dimensions, while the lower bound for the OPE-squared coefficients was set to 0. Additionally, the dimensions of the first 10 long operators were fixed completely according to the results of the QSC. 

Our results are plotted in Figure~\ref{fig:cons} against the background of the rigorous allowed regions in Figure~10 of Ref.~\cite{Cavaglia:2022qpg}. A partial list of the specific numbers with statistical errors appears in Table~\ref{tab:res2} and the full list in Table~\ref{tab:resall} of Appendix~\ref{fullresults}. 

We observe that the presence of the integral constraints significantly narrows down the statistical errors for the first OPE-coefficient squared, and enhances the accuracy of our statistical IPOPT runs, which align closely with the findings of \cite{Cavaglia:2022qpg}. This alignment is particularly noticeable for lower values of $g$. At higher values of $g$ the IPOPT results for $C_2^2$ and $C_3^2$ become less accurate, while the linear-functional method results become sharper. We believe this is due to the near-degeneracy of the corresponding operators, which makes the search more demanding. We observed that by increasing the number of parallel agents there is improvement in these numbers. 

\begin{table}
\centering
 {\small   \begin{tabular}{|l|c|c|c|c|}
 \hline
Method & $g$ & $C_1^2$ & $C_2^2$ & $C_3^2$ \\
\hline\hline
\cite{Cavaglia:2022qpg}  & $0.2$ & $0.065679029 \pm 6.95 \cdot 10^{-7}$ & $0.09452 \pm 7.25 \cdot 10^{-3}$ & $0.1101 \pm 1.27 \cdot 10^{-2}$ \\
IPOPT & $0.2$ & $0.06567873 \pm 1.55 \cdot 10^{-7}$ & $0.09683 \pm 1.41 \cdot 10^{-3}$ & $0.1063 \pm  2.42 \cdot 10^{-3}$ \\
 \hline\hline
\cite{Cavaglia:2022qpg}  & $0.4$ & $0.16838882 \pm 1.29 \cdot 10^{-6}$ & $0.06925 \pm 2.80 \cdot 10^{-3}$ & $0.13196 \pm 7.16 \cdot 10^{-3}$ \\
IPOPT & $0.4$ & $0.16838814 \pm 6.13 \cdot 10^{-7}$ & $0.07010 \pm 1.06 \cdot 10^{-3}$ & $0.13026 \pm 2.58 \cdot 10^{-3}$ \\
 \hline\hline
\cite{Cavaglia:2022qpg}  & $0.6$ & $0.233041731 \pm 4.49 \cdot 10^{-7}$ & $0.05246 \pm 1.47 \cdot 10^{-3}$ & $0.14546 \pm 2.99 \cdot 10^{-3}$ \\
IPOPT & $0.6$ & $0.233041064 \pm 8.18 \cdot 10^{-7}$ & $0.05347 \pm 1.30 \cdot 10^{-3}$ & $0.14376 \pm 2.37 \cdot 10^{-3}$ \\
 \hline\hline
\cite{Cavaglia:2022qpg}  & $0.8$ & $0.270286735 \pm 1.32 \cdot 10^{-7}$ & $0.044285 \pm 7.18 \cdot 10^{-4}$ & $0.14798 \pm 1.17 \cdot 10^{-3}$ \\
IPOPT & $0.8$ & $0.270286201 \pm 8.53 \cdot 10^{-7}$ & $0.045597 \pm  1.60 \cdot 10^{-3}$ & $0.14607 \pm 2.27 \cdot 10^{-3}$ \\
 \hline\hline
\cite{Cavaglia:2022qpg}  & $1.0$ & $0.294014873 \pm 4.88 \cdot 10^{-8}$ & $0.039788 \pm 4.10 \cdot 10^{-4}$ & $0.146757 \pm 5.82 \cdot 10^{-4}$ \\
IPOPT & $1.0$ & $0.294014228 \pm 6.77 \cdot 10^{-7}$ & $0.041832 \pm  1.86 \cdot 10^{-3}$ & $0.144100 \pm 2.39 \cdot 10^{-3}$ \\
 \hline
\end{tabular}}
\caption{\footnotesize{Partial list of results (for $g\in [0.2,1]$) from IPOPT runs with two integral constraints and comparison with \cite{Cavaglia:2022qpg}. The errors for \cite{Cavaglia:2022qpg} encode the rigorous upper and lower bounds about the indicated mean. The errors for IPOPT encode one standard deviation around the statistical reward-weighted average.}}
\label{tab:res2}
\end{table}

One might wonder whether our approximation scheme, which includes the tail evaluated in the free limit, remains valid for all values of $g$. We have checked this point by performing cursory runs, where besides fixing the dimensions of the first 10 long operators to the results of the QSC, we also fixed the values of $C_1^2, C_2^2$ to the values of \cite{Cavaglia:2022qpg}. IPOPT then recovered, for all $g$, the value of $C_3^2$ in \cite{Cavaglia:2022qpg} with at least third decimal point accuracy (and often fifth decimal point). We expect that full-fledged statistical runs would improve this even further. This provides favourable evidence that our (soft) tail approximation scheme works well in this specific problem, for a large region of parameter space from weak to strong coupling. 

Moreover, at $g=1$ we computed the sum $C_3^2 + 1.13 C_2^2$, which is expected to come in at $0.19$ from \cite{Cavaglia:2021bnz}, and found:
    \bea
        \label{sum_check_2}
        \textrm{IPOPT with constraints}&:&~~ C_3^2 + 1.13 C_2^2=0.19137\;, \\
        \textrm{Cavagli\`a et al. [2]}&:&~~ C_3^2 + 1.13 C_2^2=0.19171
        ~.
    \eea
We also computed the sum $C_2^2 + C_3^2$ of our IPOPT results for all values of $g$ (up to $g=4$) and found it to agree with \cite{Cavaglia:2022qpg} to at least three decimal points. This is further evidence for the validity of the tail approximation.

\section{Analysis and Discussion}
\label{comparison}

We would now like to discuss some of the most informative features of the approximate solutions of Sections~\ref{wointegral}, \ref{wintegral}. These properties are not apparent from the table and figure representation of the results for the first three OPE-coefficients squared. First, we will comment on the OPE coefficients of higher excited states predicted by our searches. Second, we will compare the performance of the SAC and IPOPT algorithms. In particular, we would like to address the questions: 

\vspace{0.3cm}
\noindent
{\it What have we learnt about non-convex optimisers in the context of our truncation schemes? Is Reinforcement Learning a useful tool for future studies?}

\subsection{Higher CFT Data}

We remind the reader that, in addition to the first three lowest-lying operators in the $J=1$ and $J=2$ families, our searches also had the scaling dimensions of 6 operators in the $J=3$ family and the leading operator in the $J=4$ family fixed using the QSC.
These operators acquire anomalous dimensions and their scaling dimensions can cross with other operators as functions of the coupling. The mixing of contributions from different families in our crossing equations prevents us from extracting clear results for individual operators. However, this mixing is minimal, or altogether absent, for the 4 lowest-lying operators in the $J=3$ family; their scaling dimensions (labelled  $\Delta_4, \Delta_5, \Delta_6, \Delta_8$ in the language of \cite{Cavaglia:2022qpg}) are tracked with the QSC. As a preliminary result, we have included in Table \ref{Tab_furtherfour} of Appendix \ref{fullresults} the corresponding values of the OPE-coefficients squared  $C_4^2, C_5^2, C_6^2, C_8^2$ for $g\in [0.2,1]$, obtained by independently using: SAC without integral constraints, IPOPT without and with integral constraints. These are the same runs already reported with $N_{der}=260$.

We observe that the statistical errors are now more significant, which aligns with the observations of \cite{Cavaglia:2022qpg}. Setting this aside, the values of all three methods are close, giving some confidence that they are in the neighbourhood of the exact result. Comparing with the unpublished rigorous bounds of the authors of Ref.\ \cite{Cavaglia:2022qpg}\footnote{We would like to thank the authors of \cite{Cavaglia:2021bnz,Cavaglia:2022qpg} for communication on this point.} supports the same conclusion.

It is interesting to ask how our results compare with known expectations at strong coupling. Before delving into the numbers, we must address an issue that affects our data at strong coupling. Throughout the whole range of $g$ values that we explored, both SAC and IPOPT have opted to keep the leading $J=5$ scaling dimensions close to their weak coupling values. At strong coupling (specifically $g=4$) this puts the scaling dimensions of some operators in the $J=5$ family close to the scaling dimensions of the nearly-degenerate $J=3$ operators and obscures the interpretation of our results. This effect is more pronounced in the IPOPT runs. In the SAC runs, only two operators are low enough to be nearly-degenerate with the $J=3$ operators. We expect that this issue can be remedied by using additional information from the QSC for the leading operator in the $J=5$ family, using an appropriate modification of the method recently developed in  \cite{Gromov:2023hzc} along the lines of \cite{Cavaglia:2021bnz}.

\begin{table}
\centering
 {\small   \begin{tabular}{|l|c|c|c|c|c|}
 \hline
Method &  $C_4^2 \cdot 10^3$ & $C_5^2\cdot 10^3$ & $C_6^2\cdot 10^3$ & $C_8^2\cdot 10^3$ & {\rm other} $\cdot\, 10^3$ \\
\hline\hline
IPOPT + cons & $5.57 \pm 5.83$  & $3.74 \pm 2.35$ & $3.76 \pm 2.31$ & $4.12 \pm 4.23$ & $11.18 \pm 15.44$\\
IPOPT  & $3.74 \pm 0.23$  & $3.58 \pm 0.21$    & $3.63 \pm 0.22$  & $3.39 \pm 0.20$  & $17.58 \pm 0.93 $ \\
SAC  & $7.50 \pm 3.30$ & $3.17 \pm 1.85$ & $3.96 \pm 2.67 $  & $2.68 \pm 2.16$ &  $10.74 \pm 5.14$ \\
 \hline
\end{tabular}}
\caption{\footnotesize{Preliminary results for the OPE-coefficients squared of the operators with scaling dimensions $\Delta_4, \Delta_5, \Delta_6, \Delta_8$ at $g = 4$ from the searches of Sections~\ref{wointegral} and \ref{wintegral}. The `other' contributions come from effective operators of the $J=5$ family that have comparable scaling dimensions.}}
\label{tab:res3}
\end{table}

In Table \ref{tab:res3} we present the results of our three runs at $g=4$. In the column `other' we have included the OPE-coefficient squared contribution of operators in the $J=5$ family with scaling dimensions close to the $J=3$ dimensions of interest.\footnote{At $g=4$ the nearly-degenerate operators of interest in the $J=3$ family have scaling dimensions $\Delta_4=5.504295213$, $\Delta_5=5.521481452$, $\Delta_6=5.516492991$, $\Delta_8=5.539940361$. In SAC the two interfering $J=5$ operators come out at $\Delta=5.226 \pm 0.075$ with $C^2 = 0.00742 \pm 2.88\cdot 10^{-3}$ and $\Delta=5.508 \pm 0.078$ with $C^2=0.00332 \pm 2.26\cdot 10^{-3}$. The next $J=5$ operator, which was not included in Table \ref{tab:res3}, has $\Delta=5.875 \pm 0.072$. For IPOPT the $J=5$ effective operators are more densely spread around $\Delta = 5.5$. In `other' we included $J=5$ operators within the band $\Delta \in [ 5.25, 5.7]$, which involved 6/7 operators for the constrained/unconstrained search respectively.} For comparison, Ref.\ \cite{Cavaglia:2022qpg} reports the upper bounds
\begin{equation}
    \label{newpredaa}
    C_4^2 < 0.0079~, ~~ C_5^2 < 0.0123
    ~.
\end{equation}
In addition, Ref.\ \cite{Ferrero:2021bsb} has computed the strong coupling limit of the total OPE-coefficient squared of the four $J=3$ degenerate operators at $10/429 \simeq 0.023$. Adding up the contributions in Table \ref{tab:res3} we obtain:
\bea
\label{newpredab}
{\rm IPOPT~with~constraints} &:& 0.028   \pm 0.030 \;,\cr
{\rm IPOPT~w/o~constraints} &:&  0.032 \pm 0.002 \;,\cr
{\rm SAC} &:& ~ 0.028 \pm 0.015
~.
\eea
At the current stage we do not want to read too much into these numbers, as the statistical errors are significant, but they appear to indicate that our approach is on the right track. We believe that with further improvements, such as fixing the dimension of the first $J=5$ operator from the QSC,  one will eventually be able to obtain more accurate and reliable results for these CFT data as well.

\subsection{The Unreasonable Effectiveness of the SAC Average}

Our results show that the (reward-weighted) average of the SAC runs is particularly accurate, and frequently much closer to the actual result (compared to that of the maximum-reward agent in the population). If SAC can efficiently locate a basin of attraction, as already anticipated and partially observed in \cite{Kantor:2022epi}, then perhaps this is a natural expectation. However, whether and how this actually happens is not at all obvious for several reasons. Most notably, unlike other typical (deterministic or stochastic) optimisation algorithms, where one observes a high-reward-driven distribution of configurations exploiting the micro-structure of the search landscape, in SAC individual agents are comparatively reward-underachievers. The reward of the average configuration is not remarkable either. Moreover, since we have been running a tiny population of 200 parallel agents---when IPOPT was operated with $4 \times 10^8$ agents---one may also question the quality of the statistics we obtained.   

\begin{figure}
\begin{floatrow}
\ffigbox{%
\hspace{-0.55cm}
\includegraphics[width=8.5cm]{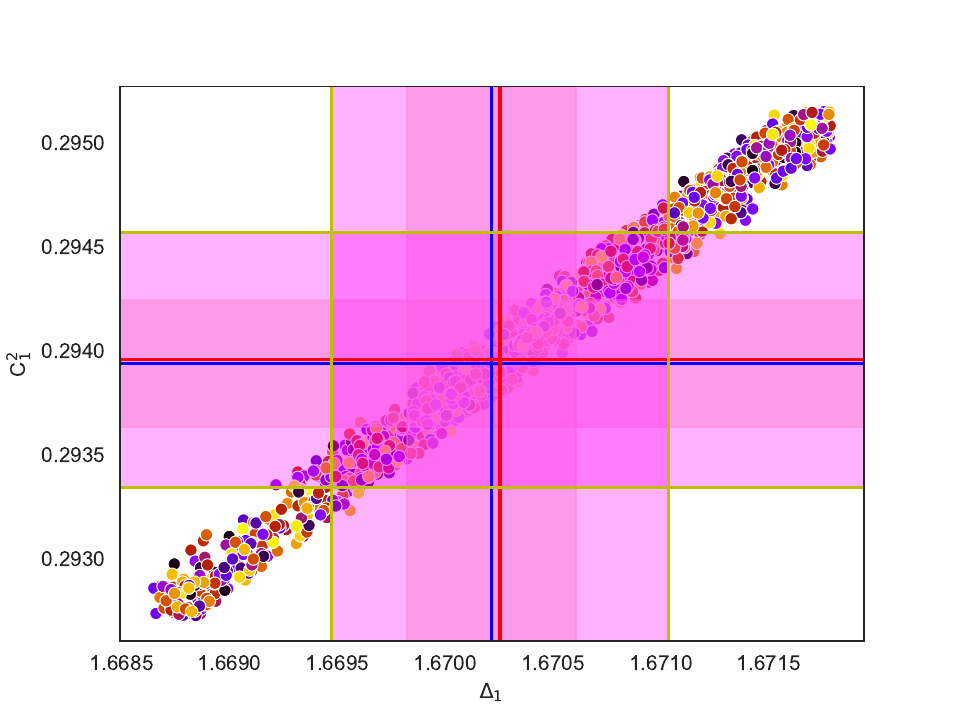}\\
\hspace{-0.55cm}
\includegraphics[width=8.5cm]{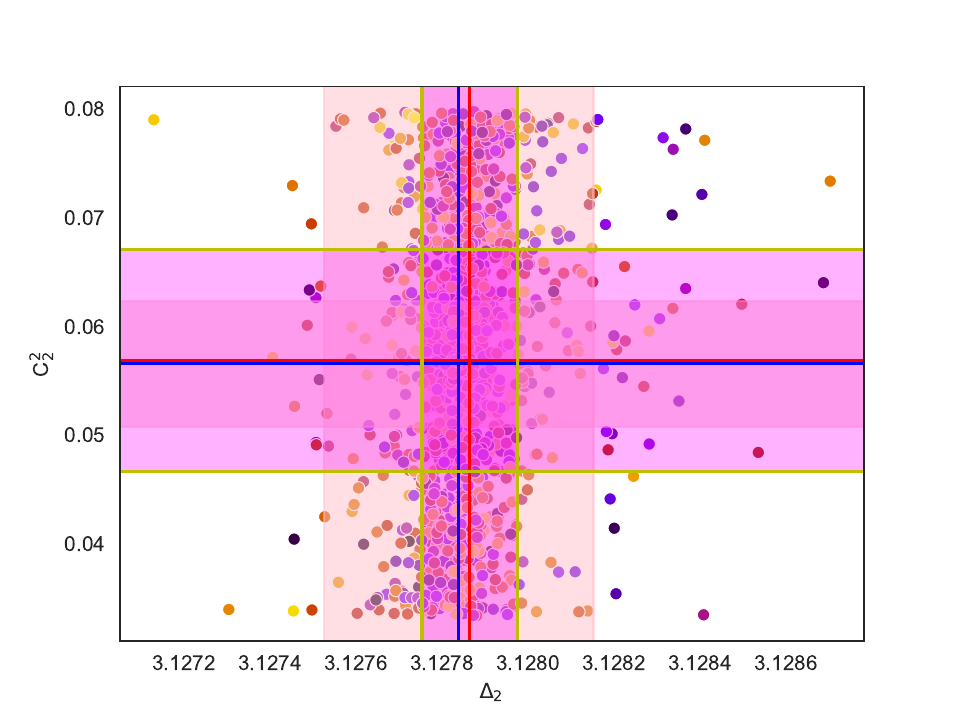}\\
\hspace{-0.55cm}
\includegraphics[width=8.5cm]{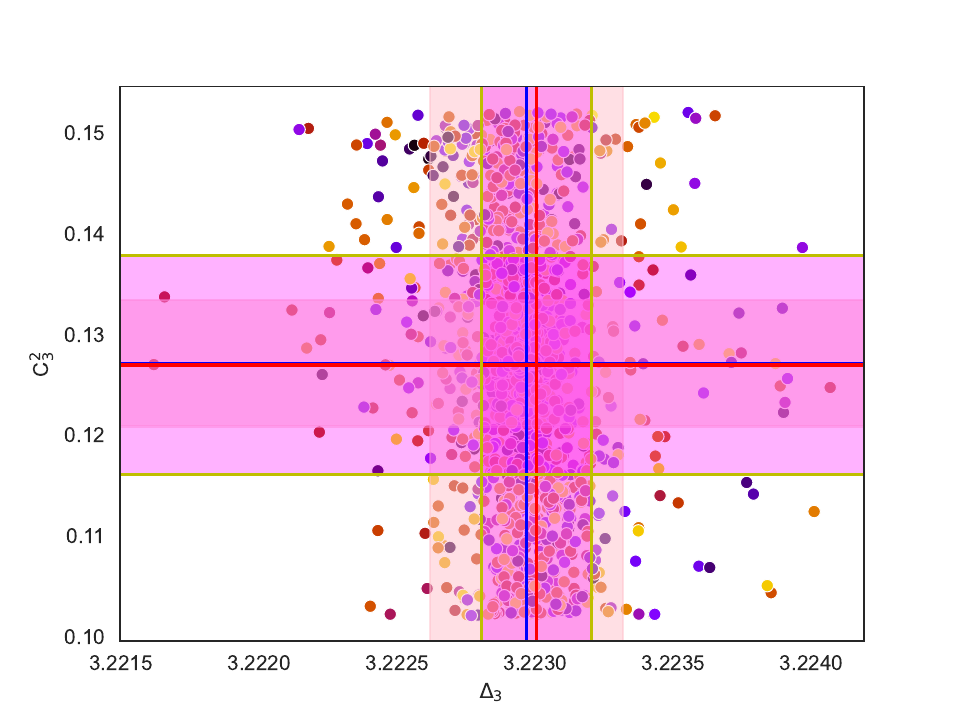}
}{%
\caption{\footnotesize{Plots of SAC and IPOPT results with $N_{der}=260$ w/o integral constraints at $g=1$. For SAC only the average (blue lines) and the $1\sigma$ regions (pink) appear. For IPOPT, we plot the average (red lines), the $1\sigma$ region (magenta) and the results of the best run for each of the $2k$ jobs on the HPC cluster (points).}}
\label{fig:operator1}
}\hspace{-0.2cm}
\capbtabbox{%
  \begin{tabular}{|l|c|c|} 
  \hline
  \footnotesize{Algorithm} & $\Delta_1$ & $C_1^2$ \\ 
  \hline\hline
  \footnotesize{QSC} & \scriptsize{1.670227842} &  \\
  \hline
  \footnotesize{IPOPT$_{260}$} & \scriptsize{$1.6702536 \pm 7.80\cdot 10^{-4}$} & \scriptsize{$0.2939600 \pm 6.12\cdot 10^{-4}$} \\
  \hline
  \footnotesize{IPOPT$_{700}$} & \scriptsize{$1.6702387 \pm 8.05\cdot 10^{-4}$} & \scriptsize{$0.2939429 \pm 6.34\cdot 10^{-4}$} \\ 
  \hline
  \footnotesize{SAC$_{260}$} & \scriptsize{$1.6702139  \pm 3.92\cdot 10^{-4}$} & \scriptsize{$0.2939411 \pm 3.03\cdot 10^{-4}$}\\
  \hline
  \multicolumn{3}{c}{} \\
  \multicolumn{3}{c}{} \\
  \multicolumn{3}{c}{} \\
  \multicolumn{3}{c}{} \\
  \multicolumn{3}{c}{} \\
  \hline
  \footnotesize{Algorithm} & $\Delta_2$ & $C_2^2$ \\ 
  \hline\hline
  \footnotesize{QSC} & \scriptsize{3.127846278} &  \\
  \hline
  \footnotesize{IPOPT$_{260}$} & \scriptsize{$3.1278644 \pm 1.11\cdot 10^{-4}$} & \scriptsize{$0.0569002 \pm 1.02\cdot 10^{-2}$} \\
  \hline
  \footnotesize{IPOPT$_{700}$} & \scriptsize{$3.1278716 \pm 1.41\cdot 10^{-4}$} & \scriptsize{$0.0582812 \pm 1.00\cdot 10^{-2}$} \\ 
  \hline
  \footnotesize{SAC$_{260}$} & \scriptsize{$3.1278389 \pm 3.13\cdot 10^{-4}$} & \scriptsize{$0.0565833 \pm 5.81\cdot 10^{-3}$}\\
  \hline 
  \multicolumn{3}{c}{} \\
  \multicolumn{3}{c}{} \\
  \multicolumn{3}{c}{} \\
  \multicolumn{3}{c}{} \\
  \multicolumn{3}{c}{} \\
  \hline 
  \footnotesize{Algorithm} & $\Delta_3$ & $C_3^2$ \\ 
  \hline\hline
  \footnotesize{QSC} & \scriptsize{3.222893829} &  \\
  \hline
  \footnotesize{IPOPT$_{260}$} & \scriptsize{$3.2230032 \pm 2.00\cdot 10^{-4}$} & \scriptsize{$0.1269864 \pm 1.09\cdot 10^{-2}$} \\
  \hline
  \footnotesize{IPOPT$_{700}$} & \scriptsize{$3.2229707 \pm 2.86\cdot 10^{-4}$} & \scriptsize{$0.1254095 \pm 1.05\cdot 10^{-2}$} \\ 
  \hline
  \footnotesize{SAC$_{260}$} & \scriptsize{$3.2229662 \pm 3.49\cdot 10^{-4}$} & \scriptsize{$0.1271357 \pm 6.26\cdot 10^{-3}$}\\
  \hline
  \end{tabular}
}{%
  \caption{\footnotesize{The average and $1\sigma$ values of the plotted CFT data for the first three long operators. The subscripts in SAC and IPOPT denote the number of maximum derivatives used. The row of each table also includes the QSC value of the corresponding scaling dimension.}}%
  \label{tab:ipopt_on_sac}
  \vspace{2.0cm}
}
\end{floatrow}
\end{figure}

In order to test the quality of the SAC average, and whether SAC was able to identify a genuine basin of attraction, we performed the following exercise at the end of all of our 200-agent SAC runs. We set a search box around the SAC average, $\overline{\rm SAC}$, with bounds $[{\overline{\rm SAC}}_i - 4 \sigma_i, {\overline{\rm SAC}}_i + 4 \sigma_i]$. The index $i$ denotes the $i$th CFT datum and $\sigma_i$ its corresponding 1$\sigma$ uncertainty in the SAC runs. Inside this box we ran $2\times 10^8$ IPOPT agents (subdivided into 2k groups each with 100k population). We repeated these `IPOPT-on-SAC' runs for all values of $g$. The results at $g=1$ are plotted in Figure~\ref{fig:operator1}. In Table \ref{tab:ipopt_on_sac} we present the corresponding values of the reward-weighted averages for the SAC and IPOPT runs with $N_{der}=260$, as well as results from IPOPT runs with $N_{der}=700$, which are not plotted in Figure~\ref{fig:operator1}. For reference we also included the QSC values of the scaling dimensions.

For all values of $g$, we observed the following features. First, the algorithms optimise the scaling dimensions in the vicinity of the QSC values. This is a satisfying minimal check of the method against the QSC expectations. Second, increasing the number of derivatives from 260 to 700 in the IPOPT runs does not appear to yield any significant improvements. Third, and most important, the IPOPT averages reproduce consistently and with great accuracy the SAC averages despite the spread that the IPOPT agents exhibit.  It is truly striking that the SAC runs with only 200 agents and relatively low reward have managed to capture well a local basin of attraction. For the first operator, we also notice an intriguing feature of Figure~\ref{fig:operator1}: the IPOPT results are arranged linearly along the diagonal on the $(\Delta_1, C_1^2)$ plane. We have observed this configuration at all values of $g$, but do not have a clear explanation for it.

For the $g=1$ results in Figure~\ref{fig:operator1} and Table \ref{tab:ipopt_on_sac}, we obtained the following rewards: 
\bea
\label{rewards}
200~{\rm SAC_{260}~agents}&:& \min = 1.76 \cdot 10^{4}~, ~~ {\rm median}=4.46 \cdot 10^{4}~, ~~ \max=1.82 \cdot 10^{6}
~,\nonumber\\
2{\rm k} ~{\rm IPOPT_{260}~agents}&:& \min = 2.52 \cdot 10^{3}~, ~~ {\rm median}=1.56 \cdot 10^{6}~, ~~ \max=2.66 \cdot 10^{7}
~,\nonumber\\
2{\rm k} ~{\rm IPOPT_{700}~agents}&:& \min = 1.17 \cdot 10^{3}~, ~~ {\rm median}=1.57 \cdot 10^{6}~, ~~ \max=2.22 \cdot 10^{7}
~.\nonumber
\eea
We observe similar features at all other values of $g$. Typically, the median and maximum rewards of the IPOPT-on-SAC runs are two orders of magnitude larger than those of the SAC runs. The above values of the rewards are also typical for all the IPOPT runs (independent of SAC) at all values of $g$, with or without the integral constraints. In conclusion, we notice that, as a powerful deterministic algorithm, IPOPT gives a visible enhancement of the reward with only small modifications to the average SAC configuration.

\subsection{Role of Effective Operators}\label{sec:effective}

In Section~\ref{truncation} we introduced and highlighted the significance of effective operators in truncation schemes. The presence (or absence) of higher-dimension effective operators can affect the quality of the results for the low-dimension data, and the freedom to rearrange them at high reward can affect the inherent uncertainties of the search. In that sense, it is not unreasonable to anticipate some correlation between the latter and the size of allowed regions in the linear-functional method. The results in Figure~\ref{fig:nocons} appear to support this expectation. 

\begin{figure}[t!]
\centering
\includegraphics[width=15cm]{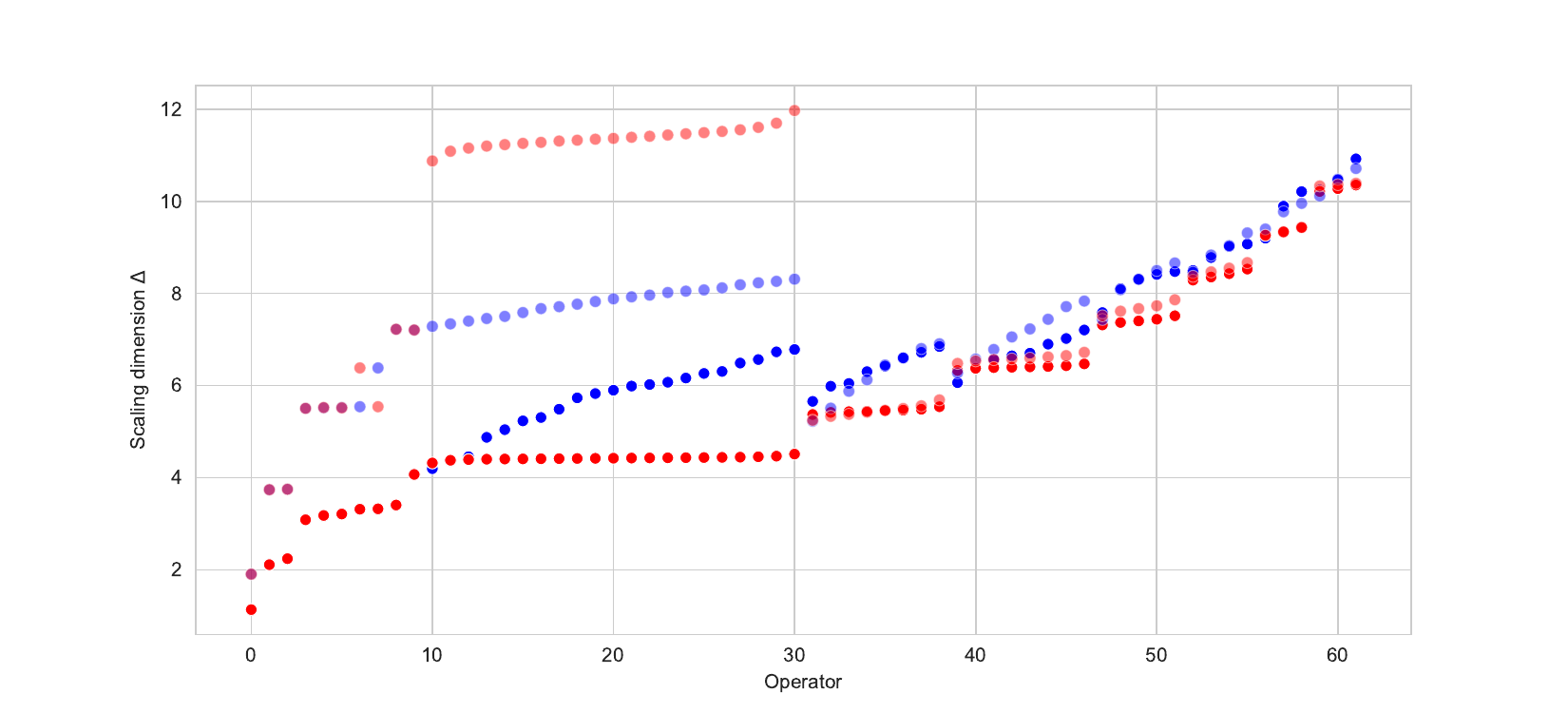}\\
\includegraphics[width=15cm]{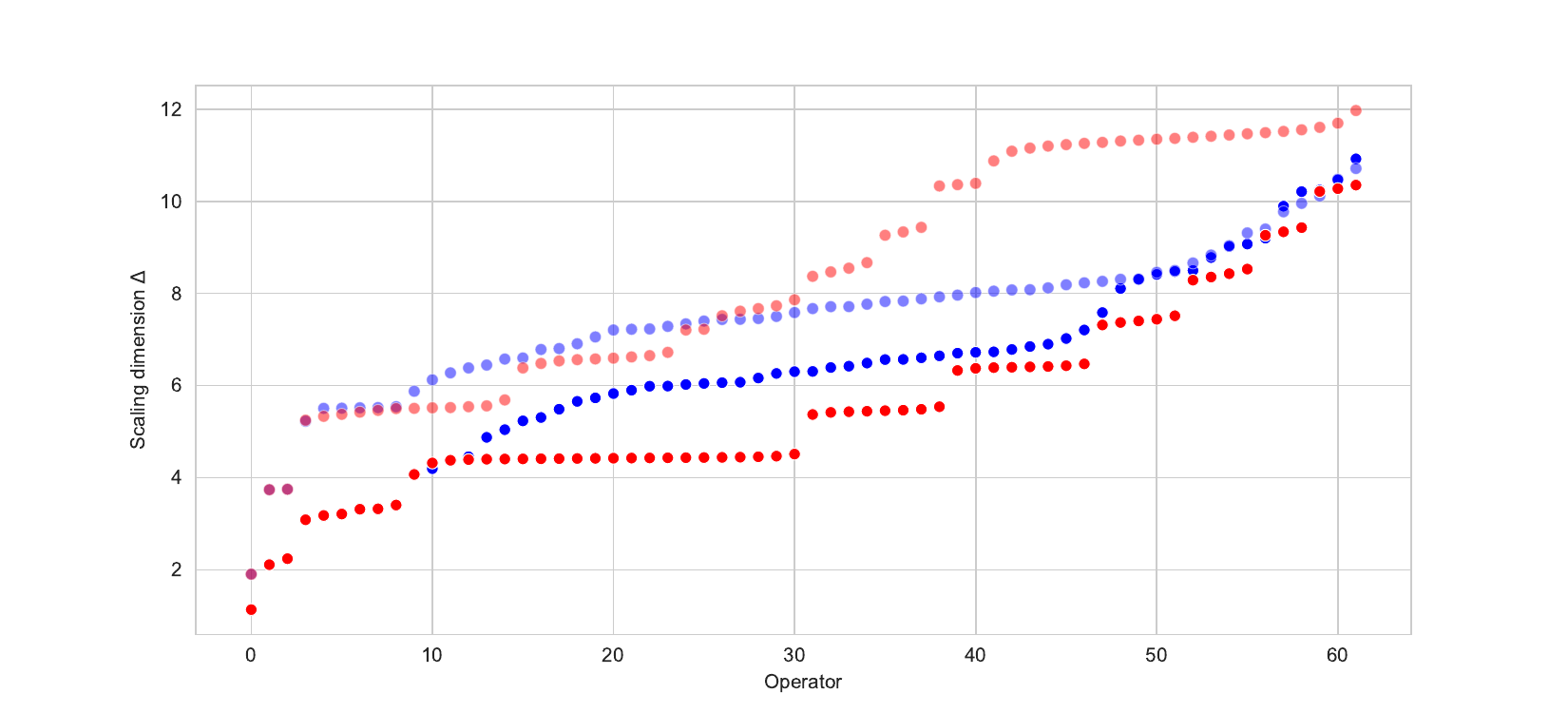}
\caption{\footnotesize{The scaling dimensions of the operators in our truncation at $g=0.2$ and $g=4$. On the $x$-axis different integer values parametrise different operators. In the upper plot the operators are ordered separately within their respective $J$ family. In the lower plot the operators are ordered globally in ascending scaling dimension. The blue dots denote SAC results: dark blue for $g=0.2$ and light blue for $g=4$. The red dots denote IPOPT results with integral constraints: dark red for $g=0.2$ and light red for $g=4$. We did not include the results of IPOPT without integral constraints, as they are very similar to the data including the constraints. The results of both SAC and IPOPT for the first 10 operators overlap.}}
\label{fig:spectrum}
\end{figure}

In the same context, it is interesting to ask how different algorithms manipulate the effective operators and, correspondingly, how they learn the landscape onto which they are optimising. In Figure~\ref{fig:spectrum} we present the scaling dimensions of all 62 operators in our truncation as obtained by SAC (dark blue for $g=0.2$ and light blue for $g=4$) and IPOPT with the integral constraints (dark red for $g=0.2$ and light red for $g=4$). Both plots contain the same information with different orderings of the operators. We observe the following features.

From the top plot in Figure~\ref{fig:spectrum} we notice that both SAC (in blue) and IPOPT (in red) have chosen to minimally vary the scaling dimensions of the $J \geq 5$ families of operators from $g=0.2$ to $g=4$. The main variation occurs for the 22 operators of the $J=4$ family and is more dramatic in the case of IPOPT. There is another significant difference between the two spectra in Figure~\ref{fig:spectrum}. IPOPT exhibits a clear tendency (at all values of $g$) to keep operators within the same family nearly-degenerate. In this manner, it effectively reduces the number of active operators in the truncation. This was a feature that was also present in other non-SAC algorithms. In contrast, SAC prefers to keep the operators more distinct, effectively smearing them across scaling dimensions, as is apparent in the bottom plot of Figure~\ref{fig:spectrum}. It would be interesting to understand why this occurs, whether it is linked to Reinforcement Learning mechanisms, and if it contains some significance that can be used constructively in future searches.  

We note in passing that, in addition to the runs reported in this paper, we also performed SAC and IPOPT runs by fixing only 9 scaling dimensions that did not include the first of the 22 operators in the $J=4$ family. In those results, the 22 $J=4$ operators received much smaller anomalous dimensions yielding qualitatively worse results both for SAC and IPOPT. It may be useful to explore if exact QSC information for the scaling dimension of the lowest $J=5$ operator, obtainable from the recent work \cite{Gromov:2023hzc}, can similarly lead to further improvements to the results presented here. Naively viewing $J$ as an analogue of spin in higher dimensions, it is tempting to consider whether this is analogous to fixing the leading Regge trajectory. That would lead to an interesting hybrid numerical bootstrap analysis in higher-dimensional CFTs, capitalising on external information from analytics, e.g.\ from OPE-inversion formulae methods. 

\subsection{On the Choice of Optimisation Algorithms}

Over the course of this work we compared the performance of several different algorithms---some deterministic, some stochastic. We observed that an algorithm like IPOPT was very efficient and could be deployed with $4\times 10^8$ agents, producing very quickly (within minutes per 100k population grouping) accurate results. In contrast, SAC produced less accurate results with lower rewards in runs that typically took 12 to 24 hours per agent, depending on the chosen scheduling for the quenching of the search-window sizes. One could therefore ask whether SAC is a useful non-convex optimisation tool within the conformal bootstrap, and whether there is sufficient motivation to explore even more advanced Machine Learning or Reinforcement Learning algorithms in this context. Based on our current understanding, we would like to argue in the affirmative for the following reasons.

In a typical study of a (truncated) crossing symmetry condition, one formulates a corresponding cost function that is then minimised by varying scaling dimensions and OPE-coefficients squared. The complicated landscape of minima may contain multiple local minimum configurations, some of which correspond to different physical theories. Therefore, when optimising, one is not necessarily looking for the global minimum. Moreover, in attempting to discover a specific local minimum one needs the ability to perform a local search without having specific information about the bounds of the search. As a result, typical optimisation algorithms that are efficient in global searches or are efficient in local searches but require specific bounds, like Interior Point methods, Stochastic Gradient Descent and others, are not in general appropriate tools. This feature was not particularly pronounced in the 1D problem that we analysed in this paper (at least for some CFT data), but it is generally an important issue. For example, in the context of the 6D (2,0) superconformal bootstrap that was analysed in \cite{Kantor:2022epi}, there are at least two minima of physical interest corresponding to the $A$ and $D$ series CFTs. In that case, setting up a global search (or a local search with prescribed bounds) is not an efficient approach. On the contrary, Markov-chain algorithms can be a useful alternative, because one can drop agents within a region of interest, and the algorithm will dynamically decide how to explore the local search landscape. SAC (and other similar stochastic algorithms) play this game especially well. SAC is particularly appealing within this class, because it is built upon a process of policy optimisation, which adapts its dynamics on the given landscape. Indeed, we have already highlighted several features within this section that support the use of algorithms like SAC. These features include the effectiveness of the average, the qualitative features of the solution and the smooth evolution across parameter space. The collection of these observations also motivates looking for improvements to our Reinforcement Learning approach, ranging from refining our SAC implementation, as well upgrading the algorithm itself by deploying Constrained and/or Multi Agent Reinforcement Learning algorithms, where the agents collaborate to achieve higher rewards.

Irrespective of the above debate, we believe that the present work, and the inherent complexity of the physical problem we are trying to solve, strongly suggest building an arsenal of diverse algorithms. These algorithms can possess traits, which can be used either separately or in combination, to learn a particular problem from qualitatively different perspectives. It is not necessarily desirable to isolate a single algorithm based on a sole feature like speed of performance or quality of the reward. This is perhaps a philosophy that departs from the standard perspective in many optimisation problems. We have updated \href{https://github.com/vniarchos/BootSTOP}{BootSTOP} to make more algorithms directly accessible to the bootstrap practitioner in service to this philosophy.

\section{Outlook}
\label{outlook}

In this paper we studied the four-point function bootstrap in the 1D CFT of the $\frac{1}{2}$-BPS straight Maldacena-Wilson line in 4D $\NN=4$ SYM theory. Importing information for the scaling dimensions of 10 long operators from the Quantum Spectral Curve \cite{Gromov:2013pga,Gromov:2014caa}, we analysed the crossing symmetry conditions with or without the inclusion of two additional sum rules arising from integrated correlation functions \cite{Drukker:2022pxk,Cavaglia:2022yvv}. Unlike Refs.\ \cite{Cavaglia:2021bnz,Cavaglia:2022qpg}, which employed linear-functional methods, we introduced an improved truncation scheme with a tail approximation that requires the solution of a non-linear, non-convex optimisation problem. We attacked this problem using stochastic (Reinforcement Learning) and deterministic (Interior Point Method) algorithms, producing numerical results for three (non-protected) OPE-coefficients squared that aligned very close to the rigorous results of \cite{Cavaglia:2021bnz,Cavaglia:2022qpg}. The main novelties and contributions of this paper are summarised at the end of Section\ \ref{summary}. Our method is relatively computationally cheap, it does not rely on positivity constraints and offers many future opportunities when combined with other analytic and rigorous numerical methods. Our Python implementation, \href{https://github.com/vniarchos/BootSTOP}{BootSTOP}, now contains a library of pre-generated conformal blocks in diverse spacetime dimensions and, besides SAC, also direct access to all the deterministic and stochastic algorithms within \href{https://github.com/esa/PyGMO2}{PyGMO} \cite{Biscani2020}.

One can envision further applications in several different directions. It would be interesting to further explore the 1D CFT of the $\frac{1}{2}$-BPS Maldacena--Wilson line aiming at the computation of more CFT data. We expect that the judicious use of additional input from the QSC and the simultaneous study of multiple correlators can lead to significant progress. Analogous progress may be possible also in other higher-dimensional CFTs. It would be interesting to revisit the 6D $\NN=(2,0)$ bootstrap \cite{Beem:2015aoa, Lemos:2021azv, Alday:2022ldo, Kantor:2022epi} and 4D $\NN=4$ superconformal bootstrap \cite{Beem:2013qxa,Beem:2016wfs,Chester:2021aun} using improved truncation schemes, and to explore in that context the improvements in accuracy and rigour that can arise with the combination of other numerical methods (e.g.\ navigator methods) and analytic methods (e.g.\ methods based on the light-cone bootstrap and the OPE-inversion formula).

\section*{Acknowledgements}

We would like to thank Andreas Stergiou for suggesting IPOPT and the PyGMO package, and for participation at the early stages of this project. We would also like to thank Alexandros Stratoudakis for collaboration on related work. We are grateful to the authors of \cite{Cavaglia:2021bnz,Cavaglia:2022qpg}, and especially Nikolay Gromov, for communication and giving us access to unpublished results. CP and PR were supported by the Science and Technology Facilities Council (STFC) Consolidated Grant ST/T000686/1 ``Amplitudes, strings $\&$ duality''. AGS acknowledges support from Pierre Andurand. MW is supported by an STFC Research Studentship. This research utilised the Apocrita HPC facility, supported by QMUL Research-IT \cite{king_thomas_2017_438045}. 

\newpage

\begin{appendix}

\section{Explicit Numerical Results}
\label{fullresults}
In this appendix we list our complete results for runs without integral constraints (both SAC and IPOPT), as well as for IPOPT runs with both integral constraints implemented. In Table~\ref{tab:resall} we list the values for the OPE-coefficients squared of the first three unprotected operators from the $J = 1,2 $ families, $C_1^2, C_2^2, C_3^2$, when $g \in [0.2,4]$ along with the results of \cite{Cavaglia:2021bnz,Cavaglia:2022qpg} for reference.

{\footnotesize
\begin{longtable}[t]{|l|c|c|c|c|}
\caption{\footnotesize{Explicit numerical results for OPE-coefficients squared $C_1^2, C_2^2, C_3^2$ from: $a)$ \cite{Cavaglia:2022qpg}, $b)$ IPOPT with two integral constraints imposed $c)$ \cite{Cavaglia:2021bnz}, $d)$ IPOPT with no integral constraints imposed and $e)$ SAC with no integral constraints imposed. The errors for $a)$ and $c)$ encode the rigorous upper and lower bounds about the indicated mean. For $b)$, $d)$, $e)$ the errors encode one standard deviation from the statistical average.\label{tab:resall}}}
\endfirsthead
\caption{\footnotesize{Explicit numerical results for $C_1^2, C_2^2, C_3^2$ continued.}}
\endhead
\hline
 Method & $g$ & $C_1^2$ & $C_2^2$ & $C_3^2$ \\
\hline\hline
\cite{Cavaglia:2022qpg} & $0.2$ & $0.06567902 \pm 6.95 \cdot 10^{-7}$ & $0.09452 \pm 7.25 \cdot 10^{-3}$ & $0.1101 \pm 1.27 \cdot 10^{-2}$ \\
IPOPT w/ cons & $0.2$ & $0.06567873 \pm 1.55 \cdot 10^{-7}$ & $0.09683 \pm 1.41 \cdot 10^{-3}$ & $0.1063 \pm  2.42 \cdot 10^{-3}$ \\
\hline \cite{Cavaglia:2021bnz} & $0.2$ & $0.0663 \pm 1.9 \cdot 10^{-3}$ &  &  \\
IPOPT w/o  cons & $0.2$ & $0.06607342 \pm 4.18 \cdot 10^{-5}$ & $0.04708 \pm 2.04 \cdot 10^{-3}$ & $0.1630 \pm 2.69 \cdot 10^{-3}$ \\
SAC w/o cons & $0.2$ & $0.06733947 \pm 1.26 \cdot 10^{-3}$ & $0.06506 \pm 1.05 \cdot 10^{-2}$ & $0.1384 \pm 1.47 \cdot 10^{-2}$ \\
 \hline\hline
\cite{Cavaglia:2022qpg} & $0.4$ & $0.16838882 \pm 1.29 \cdot 10^{-6}$ & $0.06925 \pm 2.80 \cdot 10^{-3}$ & $0.13196 \pm 7.16 \cdot 10^{-3}$ \\
IPOPT w/ cons & $0.4$ & $0.16838814 \pm 6.13 \cdot 10^{-7}$ & $0.07010 \pm 1.06 \cdot 10^{-3}$ & $0.13026 \pm 2.58 \cdot 10^{-3}$ \\
\hline \cite{Cavaglia:2021bnz} & $0.4$ & $0.1684 \pm 1.9 \cdot 10^{-3}$ &  &  \\
IPOPT w/o  cons & $0.4$ & $0.16944584 \pm 8.35 \cdot 10^{-5}$ & $0.02659 \pm 3.45 \cdot 10^{-3}$ & $0.17965 \pm 4.90 \cdot 10^{-3}$ \\
SAC w/o cons & $0.4$ & $0.16824002 \pm 1.00 \cdot 10^{-3}$ & $0.06380 \pm 1.37 \cdot 10^{-2}$ & $0.14198 \pm 1.80 \cdot 10^{-2}$ \\
 \hline\hline
\cite{Cavaglia:2022qpg} & $0.6$ & $0.233041731 \pm 4.49 \cdot 10^{-7}$ & $0.05246 \pm 1.47 \cdot 10^{-3}$ & $0.14546 \pm 2.99 \cdot 10^{-3}$ \\
IPOPT w/ cons & $0.6$ & $0.233041064 \pm 8.18 \cdot 10^{-7}$ & $0.05347 \pm 1.30 \cdot 10^{-3}$ & $0.14376 \pm 2.37 \cdot 10^{-3}$ \\
\hline \cite{Cavaglia:2021bnz} & $0.6$ & $0.2329 \pm 9 \cdot 10^{-4}$ &  &  \\
IPOPT w/o  cons & $0.6$ & $0.233574606 \pm 1.32 \cdot 10^{-4}$ & $0.02533 \pm 6.78 \cdot 10^{-3}$ & $0.17382 \pm 7.68 \cdot 10^{-3}$ \\
SAC w/o cons & $0.6$ & $0.232721152 \pm 3.24 \cdot 10^{-4}$ & $0.06151 \pm 5.46 \cdot 10^{-3}$ & $0.13363 \pm 6.77 \cdot 10^{-3}$ \\
 \hline\hline
\cite{Cavaglia:2022qpg} & $0.8$ & $0.270286735 \pm 1.32 \cdot 10^{-7}$ & $0.044285 \pm 7.18 \cdot 10^{-4}$ & $0.14798 \pm 1.17 \cdot 10^{-3}$ \\
IPOPT w/ cons & $0.8$ & $0.270286201 \pm 8.53 \cdot 10^{-7}$ & $0.045597 \pm  1.60 \cdot 10^{-3}$ & $0.14607 \pm 2.27 \cdot 10^{-3}$ \\
 \hline \cite{Cavaglia:2021bnz} & $0.8$ & $0.2701 \pm 5 \cdot 10^{-4}$ &  &  \\
IPOPT w/o  cons & $0.8$ & $0.270632286 \pm 6.67 \cdot 10^{-5}$ & $0.020165 \pm 6.29 \cdot 10^{-3}$ & $0.17218 \pm 7.06 \cdot 10^{-3}$ \\
SAC w/o cons & $0.8$ & $0.270121362 \pm 2.93 \cdot 10^{-4}$ & $0.05776 \pm 5.00 \cdot 10^{-3}$ & $0.13110 \pm 5.35 \cdot 10^{-3}$ \\
 \hline
 \hline
\cite{Cavaglia:2022qpg} & $1.0$ & $0.294014873 \pm 4.88 \cdot 10^{-8}$ & $0.039788 \pm 4.10 \cdot 10^{-4}$ & $0.146757 \pm 5.82 \cdot 10^{-4}$ \\
IPOPT w/ cons & $1.0$ & $0.294014228 \pm 6.77 \cdot 10^{-7}$ & $0.041832 \pm  1.86 \cdot 10^{-3}$ & $0.144100 \pm 2.39 \cdot 10^{-3}$ \\
\hline \cite{Cavaglia:2021bnz} & $1.0$ & $0.29388 \pm 2.7 \cdot 10^{-4}$ &  &  \\
IPOPT w/o  cons & $1.0$ & $0.294177967 \pm 6.79 \cdot 10^{-5}$ & $0.023344 \pm 9.64 \cdot 10^{-3}$ & $0.163302 \pm 1.04 \cdot 10^{-2}$ \\
SAC w/o cons & $1.0$ & $0.293941106 \pm 3.03 \cdot 10^{-4}$ & $0.05658 \pm 5.81 \cdot 10^{-3}$ & $0.127135 \pm 6.26 \cdot 10^{-3}$ \\
 \hline
 \newpage\hline
\cite{Cavaglia:2022qpg} & $1.2$ & $0.310433307 \pm 2.16 \cdot 10^{-8}$ & $0.036979 \pm 2.62 \cdot 10^{-4}$ & $0.144696 \pm 3.40 \cdot 10^{-4}$ \\
IPOPT w/ cons & $1.2$ & $0.310433 \pm 3.07 \cdot 10^{-7}$ & $0.038659 \pm 1.19 \cdot 10^{-3}$ & $0.142616 \pm 1.44 \cdot 10^{-3}$ \\
\hline \cite{Cavaglia:2021bnz} & $1.2$ & $0.31033 \pm 1.7 \cdot 10^{-4}$ &  &  \\
IPOPT w/o  cons & $1.2$ & $0.31050381 \pm 4.45 \cdot 10^{-5}$ & $0.026543 \pm 1.04 \cdot 10^{-2}$ & $0.155286 \pm  1.14 \cdot 10^{-2}$ \\
SAC w/o cons & $1.2$ & $0.310378551 \pm 2.11 \cdot 10^{-4}$ & $0.055333 \pm 4.36 \cdot 10^{-3}$ & $0.124209 \pm 4.74 \cdot 10^{-3}$ \\
 \hline\hline
\cite{Cavaglia:2022qpg} & $1.4$ & $0.322466863 \pm 1.08 \cdot 10^{-8}$ & $0.035063 \pm 1.79 \cdot 10^{-4}$ & $0.142594 \pm 2.20 \cdot 10^{-4}$ \\
IPOPT w/ cons & $1.4$ & $0.322466925 \pm 1.47 \cdot 10^{-7}$ & $0.035350 \pm  7.74 \cdot 10^{-4}$ & $0.142205 \pm  8.95 \cdot 10^{-4}$ \\
\hline \cite{Cavaglia:2021bnz} & $1.4$ & $0.32239 \pm 1.2 \cdot 10^{-4}$ &  &  \\
IPOPT w/o  cons & $1.4$ & $0.322494228 \pm 1.64 \cdot 10^{-5}$ & $0.026081 \pm 5.01 \cdot 10^{-3}$ & $0.152136 \pm 5.33 \cdot 10^{-3}$ \\
SAC w/o cons & $1.4$ & $0.322459863 \pm 2.73 \cdot 10^{-4}$ & $0.052775 \pm 6.01 \cdot 10^{-3}$ & $0.123285 \pm 5.96 \cdot 10^{-3}$ \\
 \hline\hline
\cite{Cavaglia:2022qpg} & $1.6$ & $0.331663291 \pm 5.97 \cdot 10^{-9}$ & $0.033675 \pm 1.30 \cdot 10^{-4}$ & $0.140664 \pm 1.52 \cdot 10^{-4}$ \\
IPOPT w/ cons & $1.6$ & $0.331663391 \pm 1.42 \cdot 10^{-7}$ & $0.033591 \pm 9.48 \cdot 10^{-4}$ & $0.140715 \pm 1.06 \cdot 10^{-3}$ \\
\hline \cite{Cavaglia:2021bnz} & $1.6$ & $0.33160 \pm 9 \cdot 10^{-5}$ &  &  \\
IPOPT w/o  cons & $1.6$ & $0.331696148 \pm 7.15 \cdot 10^{-6}$ & $0.017362 \pm 2.96 \cdot 10^{-3}$ & $0.157958 \pm 3.13 \cdot 10^{-3}$ \\
SAC w/o cons & $1.6$ & $0.331515801 \pm 2.68 \cdot 10^{-4}$ & $0.051282 \pm 3.94 \cdot 10^{-3}$ & $0.122229 \pm 3.99 \cdot 10^{-3}$ \\
 \hline
 \hline
\cite{Cavaglia:2022qpg} & $1.8$ & $0.338918478 \pm 3.53 \cdot 10^{-9}$ & $0.0326214 \pm 9.75 \cdot 10^{-5}$ & $0.138948 \pm 1.11 \cdot 10^{-4}$ \\
IPOPT w/ cons & $1.8$ & $0.338918374 \pm 1.22 \cdot 10^{-7}$ & $0.0339949 \pm  1.05 \cdot 10^{-3}$ & $0.137417 \pm 1.15 \cdot 10^{-3}$ \\
\hline \cite{Cavaglia:2021bnz} & $1.8$ & $0.33887 \pm 6 \cdot 10^{-5}$ &  &  \\
IPOPT w/o  cons & $1.8$ & $0.338952186 \pm 1.03 \cdot 10^{-5}$ & $0.0133279 \pm 4.15 \cdot 10^{-3}$ & $0.159124 \pm 4.30 \cdot 10^{-3}$ \\
SAC w/o cons & $1.8$ & $0.338753189 \pm 1.99 \cdot 10^{-4}$ & $0.0496961 \pm 3.67 \cdot 10^{-3}$ & $0.121366 \pm 3.98 \cdot 10^{-3}$ \\
 \hline\hline
\cite{Cavaglia:2022qpg} & $2.0$ & $0.344787161 \pm 2.21 \cdot 10^{-9}$ & $0.0317952 \pm 7.56 \cdot 10^{-5}$ & $0.1374382 \pm 8.44 \cdot 10^{-5}$ \\
IPOPT w/ cons & $2.0$ & $0.344786958 \pm 1.12 \cdot 10^{-7}$ & $0.0343623 \pm 1.22 \cdot 10^{-3}$ & $0.1346474 \pm 1.32 \cdot 10^{-3}$ \\
\hline \cite{Cavaglia:2021bnz} & $2.0$ & $0.34475 \pm 5 \cdot 10^{-5}$ &  &  \\
IPOPT w/o  cons & $2.0$ & $0.344812637 \pm 6.30 \cdot 10^{-6}$ & $0.0154283 \pm 4.13 \cdot 10^{-3}$ & $0.1543773 \pm4.30 \cdot 10^{-3}$ \\
SAC w/o cons & $2.0$ & $0.344739867 \pm 1.67 \cdot 10^{-4}$ & $0.0484264 \pm 3.37 \cdot 10^{-3}$ & $0.1202422 \pm 3.48 \cdot 10^{-3}$ \\
 \hline
 \hline
\cite{Cavaglia:2022qpg} & $2.2$ & $0.349631253 \pm 1.45 \cdot 10^{-9}$ & $0.0311296 \pm 6.02 \cdot 10^{-5}$ & $0.1361104 \pm 6.60 \cdot 10^{-5}$ \\
IPOPT w/ cons & $2.2$ & $0.34963113 \pm 1.16 \cdot 10^{-7}$ & $0.0332158 \pm 1.55 \cdot 10^{-3}$ & $0.1338673 \pm 1.65 \cdot 10^{-3}$ \\
\hline \cite{Cavaglia:2021bnz} & $2.2$ & $0.34960 \pm 4 \cdot 10^{-5}$ &  &  \\
IPOPT w/o  cons & $2.2$ & $0.349651805 \pm 4.90 \cdot 10^{-6}$ & $0.0147950 \pm 2.78 \cdot 10^{-3}$ & $0.1529520 \pm 2.84 \cdot 10^{-3}$ \\
SAC w/o cons & $2.2$ & $0.349528529 \pm 1.69 \cdot 10^{-4}$ & $0.04762628 \pm 3.27 \cdot 10^{-3}$ & $0.1193224 \pm 3.36 \cdot 10^{-3}$ \\
 \hline
\hline
\cite{Cavaglia:2022qpg} & $2.4$ & $0.353696925 \pm 9.9 \cdot 10^{-10}$ & $0.0305818 \pm 4.90 \cdot 10^{-5}$ & $0.1349397 \pm 5.30 \cdot 10^{-5}$ \\
IPOPT w/ cons & $2.4$ & $0.353696905 \pm 1.32 \cdot 10^{-7}$ & $0.0314842 \pm 2.08 \cdot 10^{-3}$ & $0.1339678 \pm 2.21 \cdot 10^{-3}$ \\
\hline \cite{Cavaglia:2021bnz} & $2.4$ & $0.353669 \pm 3.2 \cdot 10^{-5}$ &  &  \\
IPOPT w/o  cons & $2.4$ & $0.353713043 \pm 9.54 \cdot 10^{-6}$ & $0.0184944 \pm 4.62 \cdot 10^{-3}$ & $0.1472742 \pm 4.66 \cdot 10^{-3}$ \\
SAC w/o cons & $2.4$ & $0.353608797 \pm 1.34 \cdot 10^{-4}$ & $0.0452261 \pm 3.98 \cdot 10^{-3}$ & $0.1201249 \pm 4.05 \cdot 10^{-3}$ \\
 \hline
 \newpage
 \hline
\cite{Cavaglia:2022qpg} & $2.6$ & $0.357157434 \pm 7.0 \cdot 10^{-10}$ & $0.0301230 \pm 4.06 \cdot 10^{-5}$ & $0.1339028 \pm 4.34 \cdot 10^{-5}$ \\
IPOPT w/ cons & $2.6$ & $0.357157405 \pm 1.57 \cdot 10^{-7}$ & $0.0312744 \pm 2.82 \cdot 10^{-3}$ & $0.1326779 \pm 2.97 \cdot 10^{-3}$ \\
\hline \cite{Cavaglia:2021bnz} & $2.6$ & $0.357134 \pm 2.7 \cdot 10^{-5}$ &  &  \\
IPOPT w/o  cons & $2.6$ & $0.357170084 \pm 6.24 \cdot 10^{-6}$ & $0.0187927 \pm 3.21 \cdot 10^{-3}$ & $0.1454316 \pm 3.27 \cdot 10^{-3}$ \\
SAC w/o cons & $2.6$ & $0.357024241 \pm 1.56 \cdot 10^{-4}$ & $0.0453255 \pm 2.97 \cdot 10^{-3}$ & $0.1187001 \pm 2.93 \cdot 10^{-3}$ \\
 \hline\hline
\cite{Cavaglia:2022qpg} & $2.8$ & $0.360138240 \pm 5.0 \cdot 10^{-10}$ & $0.0297329 \pm 3.42 \cdot 10^{-5}$ & $0.1329800 \pm 3.63 \cdot 10^{-5}$ \\
IPOPT w/ cons & $2.8$ & $0.360138136 \pm 1.18 \cdot 10^{-7}$ & $0.0327207 \pm 2.68 \cdot 10^{-3}$ & $0.1298450 \pm 2.80 \cdot 10^{-3}$ \\
\hline \cite{Cavaglia:2021bnz} & $2.8$ & $0.360118 \pm 2.2 \cdot 10^{-5}$ &  &  \\
IPOPT w/o  cons & $2.8$ & $0.360146985 \pm 3.74 \cdot 10^{-6}$ & $0.0190562 \pm 3.74 \cdot 10^{-6}$ & $0.1438591 \pm 5.46 \cdot 10^{-3}$ \\
SAC w/o cons & $2.8$ & $0.360062267 \pm 1.26 \cdot 10^{-4}$ & $0.0445569 \pm 3.08 \cdot 10^{-3}$ & $0.1180638 \pm 3.08 \cdot 10^{-3}$ \\
 \hline\hline
\cite{Cavaglia:2022qpg} & $3.0$ & $0.362732415 \pm 3.7 \cdot 10^{-10}$ & $0.0293973 \pm 2.92 \cdot 10^{-5}$ & $0.1321546 \pm 3.07 \cdot 10^{-5}$ \\
IPOPT w/ cons & $3.0$ & $0.362732181 \pm 8.27 \cdot 10^{-8}$ & $0.0358479 \pm 2.28 \cdot 10^{-3}$ & $0.1254397 \pm 2.38 \cdot 10^{-3}$ \\
\hline \cite{Cavaglia:2021bnz} & $3.0$ & $0.362715 \pm 1.9 \cdot 10^{-5}$ &  &  \\
IPOPT w/o  cons & $3.0$ & $0.362739018 \pm 1.46 \cdot 10^{-6}$ & $0.0237969 \pm 2.40 \cdot 10^{-3}$ & $0.1377698 \pm 2.45 \cdot 10^{-3}$ \\
SAC w/o cons & $3.0$ & $0.362574889 \pm 1.71 \cdot 10^{-4}$ & $0.0436489 \pm 2.77 \cdot 10^{-3}$ & $0.1180333 \pm 2.88 \cdot 10^{-3}$ \\
 \hline
 \hline
\cite{Cavaglia:2022qpg} & $3.2$ & $0.365010449 \pm 2.8 \cdot 10^{-10}$ & $0.0291054 \pm 2.52 \cdot 10^{-5}$ & $0.1314126 \pm 2.64 \cdot 10^{-5}$ \\
IPOPT w/ cons & $3.2$ & $0.365010269 \pm 1.11 \cdot 10^{-7}$ & $0.0348757 \pm 3.37 \cdot 10^{-3}$ & $0.1254316 \pm 3.49 \cdot 10^{-3}$ \\
\hline \cite{Cavaglia:2021bnz} & $3.2$ & $0.364995 \pm 1.6 \cdot 10^{-5}$ &  &  \\
IPOPT w/o  cons & $3.2$ & $0.365018232 \pm 2.83 \cdot 10^{-6}$ & $0.0206917 \pm 4.75 \cdot 10^{-3}$ & $0.1398753 \pm 4.85 \cdot 10^{-3}$ \\
SAC w/o cons & $3.2$ & $0.364922694 \pm 1.43 \cdot 10^{-4}$ & $0.0425634 \pm 2.43 \cdot 10^{-3}$ & $0.1179391 \pm 2.43 \cdot 10^{-3}$ \\
 \hline\hline
\cite{Cavaglia:2022qpg} & $3.4$ & $0.367026704 \pm 2.2 \cdot 10^{-10}$ & $0.0288492 \pm 2.20 \cdot 10^{-5}$ & $0.1307425 \pm 2.30 \cdot 10^{-5}$ \\
IPOPT w/ cons & $3.4$ & $0.367026579 \pm 1.47 \cdot 10^{-7}$ & $0.0335769 \pm 4.71 \cdot 10^{-3}$ & $0.1258575 \pm 4.86 \cdot 10^{-3}$ \\
\hline \cite{Cavaglia:2021bnz} & $3.4$ & $0.367013 \pm 1.4 \cdot 10^{-5}$ &  &  \\
IPOPT w/o  cons & $3.4$ & $0.36703306 \pm 2.72 \cdot 10^{-6}$ & $0.0221385 \pm  1.63 \cdot 10^{-3}$ & $0.1374627 \pm 1.65 \cdot 10^{-3}$ \\
SAC w/o cons & $3.4$ & $0.366989521 \pm 1.28 \cdot 10^{-4}$ & $0.0413939 \pm 2.98 \cdot 10^{-3}$ & $0.1180792 \pm 2.99 \cdot 10^{-3}$ \\
 \hline
\hline
\cite{Cavaglia:2022qpg} & $3.6$ & $0.368823769 \pm 1.7 \cdot 10^{-10}$ & $0.0286224 \pm 1.94 \cdot 10^{-5}$ & $0.1301347 \pm 2.02 \cdot 10^{-5}$ \\
IPOPT w/ cons & $3.6$ & $0.368823674 \pm 2.03 \cdot 10^{-7}$ & $0.0327516 \pm 7.18 \cdot 10^{-3}$ & $0.1258821 \pm 7.38 \cdot 10^{-3}$ \\
\hline \cite{Cavaglia:2021bnz} & $3.6$ & $0.368812 \pm 1.2 \cdot 10^{-5}$ &  &  \\
IPOPT w/o  cons & $3.6$ & $0.368834396 \pm 7.83 \cdot 10^{-6}$ & $0.0193471 \pm 4.10 \cdot 10^{-3}$ & $0.1393921 \pm 4.18 \cdot 10^{-3}$ \\
SAC w/o cons & $3.6$ & $0.368792133 \pm 1.50 \cdot 10^{-4}$ & $0.0411893 \pm 3.09 \cdot 10^{-3}$ & $0.1173746 \pm 3.11 \cdot 10^{-3}$ \\
 \hline
 \hline
\cite{Cavaglia:2022qpg} & $3.8$ & $0.370435484 \pm 1.4 \cdot 10^{-10}$ & $0.0284203 \pm 1.73 \cdot 10^{-5}$ & $0.1295811 \pm 1.79 \cdot 10^{-5}$ \\
IPOPT w/ cons & $3.8$ & $0.370435381 \pm 2.20 \cdot 10^{-7}$ & $0.0335783 \pm 8.74 \cdot 10^{-3}$ & $0.1242843 \pm 8.96 \cdot 10^{-3}$ \\ 
\hline \cite{Cavaglia:2021bnz} & $3.8$ & $0.370425 \pm 1.1 \cdot 10^{-5}$ &  &  \\
IPOPT w/o  cons & $3.8$ & $0.370445095 \pm 7.19 \cdot 10^{-6}$ & $0.0180924 \pm  5.66 \cdot 10^{-3}$ & $0.1399174 \pm 5.75 \cdot 10^{-3}$ \\
SAC w/o cons & $3.8$ & $0.370403920 \pm 1.67 \cdot 10^{-4}$ & $0.0411710 \pm 2.97 \cdot 10^{-3}$ & $0.1165761 \pm 2.99 \cdot 10^{-3}$ \\
 \hline
\newpage
\hline
\cite{Cavaglia:2022qpg} & $4.0$ & $0.371889072 \pm 1.1 \cdot 10^{-10}$ & $0.0282390 \pm 1.55 \cdot 10^{-5}$ & $0.1290748 \pm 1.60 \cdot 10^{-5}$ \\
IPOPT w/ cons & $4.0$ & $0.371888949 \pm 2.07 \cdot 10^{-7}$ & $0.0343581 \pm 8.98 \cdot 10^{-3}$ & $0.1228095 \pm 9.18 \cdot 10^{-3}$ \\
\hline \cite{Cavaglia:2021bnz} & $4.0$ & $0.371880 \pm 1 \cdot 10^{-5}$ &  &  \\
IPOPT w/o  cons & $4.0$ & $0.371890446 \pm 5.79 \cdot 10^{-6}$ & $0.0230225 \pm 8.59 \cdot 10^{-4}$ & $0.1343690 \pm 8.23 \cdot 10^{-4}$ \\
SAC w/o cons & $4.0$ & $0.371940952 \pm 2.1 \cdot 10^{-4}$ & $0.0404640 \pm 2.60 \cdot 10^{-3}$ & $0.1162919 \pm 2.86 \cdot 10^{-3}$ \\
\hline
\end{longtable}}

In Table~\ref{Tab_furtherfour} we list the results of the  OPE-coefficients squared for the next four unprotected operators from the $J = 3$ family, $C_4^2, C_5^2, C_6^2, C_8^2$ when $g \in [0.2,1]$. The unpublished rigorous bounds of the authors of \cite{Cavaglia:2022qpg} support the expectation that these results are on the right track. The operators with scaling dimensions $\Delta_7, \Delta_9$ acquire large anomalous dimensions at strong coupling and are no longer almost degenerate with $\Delta_4,\Delta_5,\Delta_6,\Delta_8$. The OPE-coefficient squared values for the full list of $J = 3$ unprotected operators for $g \in [0.2,4]$ are available upon request. 

\begin{table}[h!]
\centering
{\scriptsize\begin{tabular}{|l | c | c | c| c| c|} 
\hline
Method & $g$ & $C_4^2$ & $C_5^2$ & $C_6^2$ & $C_8^2$ \\
\hline\hline
IPOPT w/ cons & $0.2$ & $0.04110 \pm 1.17 \cdot 10^{-2}$ & $0.02602 \pm 7.65 \cdot 10^{-3}$ & $0.02292 \pm 9.63 \cdot 10^{-3}$ & $0.01618 \pm 5.63 \cdot 10^{-3}$ \\
IPOPT w/o  cons & $0.2$ & $0.03257 \pm 3.60 \cdot 10^{-3}$ & $0.02464 \pm 2.03 \cdot 10^{-3}$ & $0.02251 \pm 1.91 \cdot 10^{-3}$ & $0.01609 \pm 1.19 \cdot 10^{-3}$ \\
SAC w/o cons & $0.2$ & $0.02186 \pm 2.17 \cdot 10^{-2}$ & $0.03303 \pm 2.77 \cdot 10^{-2}$ & $0.01103 \pm 1.26 \cdot 10^{-2}$ & $0.02534 \pm 1.41 \cdot 10^{-2}$ \\
 \hline\hline
IPOPT w/ cons & $0.4$ & $0.02596 \pm 1.04 \cdot 10^{-2}$ & $0.02397 \pm 1.19 \cdot 10^{-2}$ & $0.02249 \pm 1.11 \cdot 10^{-2}$ & $0.01203 \pm 6.46 \cdot 10^{-3}$ \\
IPOPT w/o  cons & $0.4$ & $0.03601 \pm 5.49 \cdot 10^{-3}$ & $0.01670 \pm 2.00 \cdot 10^{-3}$ & $0.01478 \pm 1.99 \cdot 10^{-3}$ & $0.00883 \pm 1.59 \cdot 10^{-3}$ \\
SAC w/o cons & $0.4$ & $0.02191 \pm 1.65 \cdot 10^{-2}$ & $0.01655 \pm 1.24 \cdot 10^{-2}$ & $0.01546 \pm 1.35 \cdot 10^{-2}$ & $0.01987 \pm 1.73 \cdot 10^{-2}$ \\
 \hline\hline
IPOPT w/ cons & $0.6$ & $0.01376 \pm 8.94 \cdot 10^{-3}$ & $0.02165 \pm 1.35 \cdot 10^{-2}$ & $0.01875 \pm 1.09 \cdot 10^{-2}$ & $0.01365 \pm 6.03 \cdot 10^{-3}$ \\
IPOPT w/o  cons & $0.6$ & $0.02907 \pm 5.97 \cdot 10^{-3}$ & $0.01232 \pm 1.93 \cdot 10^{-3}$ & $0.01143 \pm 1.94 \cdot 10^{-3}$ & $0.00769 \pm 2.30 \cdot 10^{-3}$ \\
SAC w/o cons & $0.6$ & $0.02211 \pm 9.74 \cdot 10^{-3}$ & $0.01587 \pm 7.10 \cdot 10^{-3}$ & $0.01923 \pm 1.08 \cdot 10^{-2}$ & $0.00809 \pm 8.86 \cdot 10^{-3}$ \\
 \hline\hline
IPOPT w/ cons & $0.8$ & $0.01083 \pm 6.21 \cdot 10^{-3}$ & $0.01914 \pm 1.00 \cdot 10^{-2}$ & $0.01228 \pm 6.44 \cdot 10^{-3}$ & $0.01735 \pm 8.83 \cdot 10^{-3}$ \\
IPOPT w/o  cons & $0.8$ & $0.02839 \pm 5.43 \cdot 10^{-3}$ & $0.01005 \pm 3.40 \cdot 10^{-3}$ & $0.01043 \pm 3.28 \cdot 10^{-3}$ & $0.00548 \pm 2.58 \cdot 10^{-3}$ \\
SAC w/o cons & $0.8$ & $0.02026 \pm 6.00 \cdot 10^{-3}$  & $0.01912 \pm 7.46 \cdot 10^{-3}$& $0.01510 \pm 6.27 \cdot 10^{-3}$ & $0.00606 \pm 6.04 \cdot 10^{-3}$ \\
 \hline\hline
IPOPT w/ cons & $1.0$ & $0.01273 \pm 6.68 \cdot 10^{-3}$ & $0.01011 \pm 7.32 \cdot 10^{-3}$ & $0.01140 \pm 8.15 \cdot 10^{-3}$ & $0.01965 \pm 9.77 \cdot 10^{-3}$ \\
IPOPT w/o  cons & $1.0$ & $0.02194 \pm 6.32 \cdot 10^{-3}$ & $0.01017 \pm 3.30 \cdot 10^{-3}$ & $0.01050 \pm 3.13 \cdot 10^{-3}$ & $0.00649 \pm 3.70 \cdot 10^{-3}$ \\
SAC w/o cons & $1.0$ & $0.01712 \pm 5.97 \cdot 10^{-3}$  & $0.01688 \pm 6.40 \cdot 10^{-3}$ & $0.01628 \pm 8.17 \cdot 10^{-3}$ & $0.00519 \pm 4.66 \cdot 10^{-3}$ \\
\hline
\end{tabular}}
\caption{\footnotesize{Explicit numerical results for OPE-coefficients squared $C_4^2, C_5^2, C_6^2, C_8^2$ from: $a)$ IPOPT with two integral constraints imposed, $b)$ IPOPT with no integral constraints imposed and $c)$ SAC with no integral constraints imposed. The errors encode one standard deviation from the statistical average.}}
\label{Tab_furtherfour}
\end{table}

\section{Flow Equations}
\label{flow}

In this appendix we present an alternative analysis for the adiabatic variation of the crossing equations, which aims to determine the infinitesimal vector of the CFT data in parameter space. The approach is similar in spirit to the flowing method of Ref.\ \cite{El-Showk:2016mxr}, except that it is not specific to extremal solutions and works within a general truncation scheme. We did not use this method to derive any of the results reported in the main part of this work, but we present it here as an idea that may be useful in future studies. Preliminary analysis shows that this method can produce high reward curves, at least within a finite region of parameter space. An interesting technical aspect of this approach is that it leads to a convex optimisation problem, which is significantly simpler compared to the non-convex problems we had to solve in the main text.

As always in this paper, we assume that we have a set of crossing equations (possibly supplemented with additional sum rules) for CFT data on a parameter space.  The parameters are collectively denoted as $\lambda$ and can be continuous or discrete. If discrete, we compute finite differences instead of derivatives. Our crossing equations take the form \eqref{truncac}, repeated here as
\begin{equation}
    \label{flowaa}
    \sum_n \CC_n(\lambda) \vec F_n(\lambda) + \vec T(\lambda) + \vec r(\lambda) = 0
    ~.
\end{equation}
We assume to have an exact or approximate solution of the CFT data at $\lambda$ and are interested in determining the solution at $\lambda + \delta \lambda$ for infinitesimal (or small finite) $\delta \lambda$.

Varying \eqref{flowaa} with respect to $\lambda$ we obtain the flow equation:
\begin{equation}
    \label{flowab}
    \sum_n \left[  \vec F_n(\lambda)\, \delta \CC_n + \CC_n(\lambda) \frac{\partial \vec F_n}{\partial \Delta_n}(\lambda)\, \delta \Delta_n \right] + \delta \vec T + \delta \vec r = 0
    ~.
\end{equation}
We know by construction $\vec r$ as a function of $\lambda$ and can approximate $\delta \vec r \simeq \frac{\partial \vec r}{\partial \lambda} \delta \lambda$. In addition, by implementing our basic approximating assumption in the main text that the tail $\vec T$ has a weak $\lambda$-dependence, we can set $\delta \vec T \simeq 0$. These assumptions lead to the simplified approximate flow equation
\begin{equation}
    \label{flowac}
    \sum_n \left[  \vec F_n(\lambda)\, \delta \CC_n + \CC_n(\lambda) \frac{\partial \vec F_n}{\partial \Delta_n}(\lambda)\, \delta \Delta_n \right] + \frac{\partial \vec r}{\partial \lambda} \delta \lambda \simeq 0
    ~,
\end{equation}
which is linear in the flow-vector unknowns $\delta \CC_n, \delta \Delta_n$. An obvious strategy for proceeding is to try and solve for these unknowns by minimising a quadratic cost function. This is a linear regression problem that can be handled with one's favourite deterministic optimiser. Unfortunately, this strategy is not optimal because it relies on the assumption that at every step one flows away from a good quality solution of the CFT data. As a result, even when starting from an initial point with knowledge of the exact solution, the errors will pile up quickly and the quality of the search will decrease. 

A more promising alternative is to consider the linearised version of the full crossing equation, not just its differential. Indeed, at $\lambda + \delta \lambda$ one could aim to solve the approximate equation
\begin{equation}
    \label{flowad}
    \sum_n \CC_n(\lambda) \vec F_n(\lambda) + \vec T(\lambda) + \vec r(\lambda) +
    \sum_n \left[  \vec F_n(\lambda)\, \delta \CC_n + \CC_n(\lambda) \frac{\partial \vec F_n}{\partial \Delta_n}(\lambda)\, \delta \Delta_n \right] + \frac{\partial \vec r}{\partial \lambda} \delta \lambda \simeq 0\;.
\end{equation}
The unknowns are again the flow-vector data $\delta \CC_n, \delta \Delta_n$ and the problem is a linear regression problem. Since we included the zeroth order part in the first three terms, this approach has a chance to self-correct as one flows away from an exact solution. 

We have noticed in preliminary applications that this adiabatic flow can be run quickly with very small step $\delta \lambda$ to produce high-reward curves, but we have not studied it systematically. At finite distance in the flow one could also combine it, or compare it, with a full non-convex optimisation run (like the ones reported in the main text). On the technical side, we noticed that a quick implementation of the linear-regression in this context was possible using the SLSQP algorithm from NLopt, which is available within PyGMO. SLSQP is a sequential quadratic programming algorithm for non-linearly constrained gradient-based optimisation. We also noticed that the near-degeneracies caused instabilities in the gradient descent which could be addressed with the Ridge-regression regularisation scheme common in Machine Learning; for a review see \cite{james2014introduction}.

\end{appendix}


\bibliography{1Dbootstrap}

\end{document}